\documentclass[letterpaper,twocolumn,10pt]{article}
\usepackage{usenix}
\usepackage{amssymb}
\usepackage{amsfonts}
\usepackage{amsthm}
\usepackage{oplotsymbl}
\usepackage{graphicx}
\usepackage{textcomp}
\usepackage{units}
\usepackage{colortbl}
\usepackage{xcolor}
\usepackage[font=small]{caption}
\usepackage{tikz}
\usepackage{pgfplots}
\usepackage[T1]{fontenc}
\usepackage{amsmath}
\usepackage{array}                
\usepackage{url}
\usepackage{multirow} 
\usepackage{subfigure}
\usepackage{bm}
\usepackage{mathtools}
\usepackage{comment}
\usepackage{multicol}
\usepackage{float}
 \usepackage{balance}
\usepackage{xspace}
\usepackage{algorithm}
\usepackage{algorithmicx}
\usepackage{algpseudocode}
\usepackage{booktabs}
\usepackage{soul}
 \usepackage[normalem]{ulem}
\usepackage{enumitem}
\usepackage{todonotes}
\usetikzlibrary{patterns}
\usepackage{amssymb}

\usetikzlibrary{positioning}
\usetikzlibrary{shapes.geometric, arrows}
\pgfplotsset{compat=1.14}

\definecolor{CiteRed}{HTML}{B22222}
\definecolor{LinkGreen}{HTML}{157F3B}

\hypersetup{colorlinks=true,citecolor=CiteRed,linkcolor=LinkGreen,%
            urlcolor=LinkGreen,filecolor=LinkGreen}
\newcommand{\dimref}[1]{\hyperref[tab:taxonomy]{\textcolor{LinkGreen}{\textbf{#1}}}}
\newcommand{\cellref}[1]{\hyperref[tab:cells]{\textcolor{LinkGreen}{#1}}}
\newcommand{\appref}[1]{\S\ref{#1}}

\usepackage[subpreambles=false]{standalone}
\usepackage{pgfplots}
\usepackage{tikz}
\pgfplotsset{compat=1.18}
\usetikzlibrary{patterns,patterns.meta,positioning,calc}
\usepgfplotslibrary{groupplots}

\definecolor{cbBlue}{RGB}{0,114,178}      
\definecolor{cbVermillion}{RGB}{213,94,0} 
\definecolor{cbGreen}{RGB}{0,158,115}     
\definecolor{cbOrange}{RGB}{230,159,0}    
\definecolor{cbSky}{RGB}{86,180,233}      
\definecolor{cbPurple}{RGB}{204,121,167}  
\definecolor{cbYellow}{RGB}{240,228,66}
\definecolor{cbGray}{RGB}{120,120,120}    
\definecolor{cbInk}{RGB}{40,40,40}

\colorlet{cBull}{cbBlue}
\colorlet{cSui}{cbVermillion}
\colorlet{cFix}{cbGreen}
\colorlet{cAttack}{cbVermillion}
\colorlet{cNeutral}{cbGray}
\colorlet{cPos}{cbBlue}      
\colorlet{cMEV}{cbOrange}    

\tikzset{
  patBull/.style   ={postaction={pattern={Lines[angle=45,distance=3pt,line width=0.5pt]},pattern color=white}},
  patSui/.style    ={postaction={pattern={Lines[angle=-45,distance=3pt,line width=0.5pt]},pattern color=white}},
  patFix/.style    ={postaction={pattern={Lines[angle=0,distance=3pt,line width=0.5pt]},pattern color=white}},
  patDefault/.style={postaction={pattern={Dots[distance=3pt,radius=0.4pt]},pattern color=white}},
  patNA/.style     ={pattern={Lines[angle=45,distance=2.2pt,line width=0.4pt]},pattern color=cbGray},
}

\pgfplotsset{
  mevaxis/.style={
    width=8.4cm, height=6.0cm,
    font=\small,
    axis line style={cbInk},
    tick style={cbInk},
    label style={cbInk},
    tick label style={cbInk},
    grid=major,
    grid style={draw=black!12},
    every axis title/.style={font=\small\bfseries, cbInk},
    legend cell align=left,
    legend style={font=\footnotesize, draw=black!25, fill=white, fill opacity=0.9,
                  text opacity=1, rounded corners=1pt},
    tick align=outside,
    line width=0.9pt,
    mark size=1.8pt,
  },
  mevbar/.style={
    mevaxis, ybar, bar width=7pt,
    ymajorgrids=true, xmajorgrids=false,
  },
}

\newcommand{\FixTB}{\textsc{Fix\-Tiebreak}}

\newcommand{\Sui}{Sui}

\definecolor{AttackFissure}{HTML}{C0392B}
\definecolor{AttackSluggish}{HTML}{2C7FB8}
\definecolor{AttackSpeculative}{HTML}{7B3294}
\pgfplotsset{
  mevBarAxis/.style={
    width=\linewidth,
    height=0.33\linewidth,
    ymin=0,
    ymajorgrids=true,
    grid style={draw=black!10},
    axis line style={draw=black!55},
    tick style={draw=black!55},
    tick label style={font=\scriptsize},
    label style={font=\small},
    title style={font=\small\bfseries},
    legend style={font=\scriptsize, draw=none, fill=none},
    bar width=7pt,
  },
}

\definecolor{AttackSilent}{HTML}{1B7837}
\tikzset{
  patFissure/.style ={postaction={pattern={Lines[angle=45,distance=3pt,line width=0.5pt]},pattern color=black!55}},
  patSpec/.style    ={postaction={pattern={Dots[distance=3pt,radius=0.5pt]},pattern color=black!55}},
  patSlug/.style    ={postaction={pattern={grid[distance=3pt,line width=0.4pt]},pattern color=black!55}},
  patSilent/.style  ={postaction={pattern={Lines[angle=-45,distance=3pt,line width=0.5pt]},pattern color=black!55}},
  patBase/.style    ={},
}
\tikzset{
  barFissure/.style={draw=AttackFissure!70!black, fill=AttackFissure!35, patFissure},
  barSpec/.style   ={draw=AttackSpeculative!70!black, fill=AttackSpeculative!35, patSpec},
  barSlug/.style   ={draw=AttackSluggish!70!black, fill=AttackSluggish!35, patSlug},
  barSilent/.style ={draw=AttackSilent!70!black, fill=AttackSilent!35, patSilent},
  barBase/.style   ={draw=black!60, fill=black!14, patBase},
}

\usepgfplotslibrary{groupplots}
\pgfplotsset{
  mevAxis/.style={
    width=0.31\linewidth,
    height=0.24\linewidth,
    ymin=0,
    ymax=100,
    ymajorgrids=true,
    grid style={draw=black!10},
    axis line style={draw=black!55},
    tick style={draw=black!55},
    tick label style={font=\scriptsize},
    label style={font=\small},
    title style={font=\small\bfseries},
    legend style={
      font=\scriptsize,
      draw=none,
      fill=none,
      legend cell align=left,
      /tikz/every even column/.append style={column sep=0.35cm},
    },
    line width=1pt,
    mark size=2.2pt,
  },
}

\definecolor{ProtoNarwhal}{HTML}{1B9E77}
\definecolor{ProtoBullshark}{HTML}{D95F02}
\definecolor{ProtoMysticeti}{HTML}{7570B3}
\definecolor{ProtoAleph}{HTML}{66A61E}
\definecolor{ProtoMahi}{HTML}{E6AB02}
\definecolor{ProtoAutobahn}{HTML}{1F78B4}

\setlist{nosep}

\makeatletter
\def\thm@space@setup{\thm@preskip=-4pt
\thm@postskip=0pt}
\makeatother

\theoremstyle{remark}
\newtheorem{adver}{{\bf A}}
\newtheorem{proto}{{\bf P}}
\newtheorem{target}{{\bf T}}
\newtheorem{deploy}{{\bf D}}

\newlength{\affwd}
\newif\ifextend
\extendtrue

\begin{document}

\title{\Large \bf Competition, Collusion, and Corruption:\\ The Spectrum of MEV Attacks on
DAG-Based BFT Consensus Protocols}

\author{%
\setlength{\parindent}{0pt}%
\settowidth{\affwd}{City University of Hong Kong}%
\begin{tabular}{@{}c@{\hspace{1.6em}}c@{\hspace{1.6em}}c@{}}
\makebox[\affwd][c]{\rm Iliya Mirzaei} & \makebox[\affwd][c]{\rm Heer Patel} & \makebox[\affwd][c]{\rm Chenyuan Wu}\\
\makebox[\affwd][c]{Stony Brook University} & \makebox[\affwd][c]{Stony Brook University} & \makebox[\affwd][c]{City University of Hong Kong}
\end{tabular}\\[0.9em]
{\rm Mohammad Javad Amiri}\\
Stony Brook University%
}

\maketitle

\begin{abstract}
Byzantine Fault-Tolerant (BFT) protocols guarantee safety and liveness despite the malicious failure of nodes. However, they do not prevent adversarial manipulation of transaction order, where the order a proposer assigns diverges from the order in which clients submitted their transactions. Exploiting this discretion for profit is known as maximal extractable value (MEV), and it is intensified in DAG-based BFT protocols, where every replica proposes blocks concurrently rather than routing transactions through a single designated proposer each round. The proliferation of MEV attacks on DAG-based BFT protocols has made the resulting landscape difficult to navigate: attacks are reported individually, on different protocols, and under different metrics, making it unclear whether two attacks differ fundamentally or merely in how they are described. This paper closes that gap by presenting an attack space for MEV on DAG-based BFT protocols, organized around four families: the adversary, the protocol, the target, and the deployment. For each family, we identify the dimensions that shape an attack's impact. Each point in the attack space fixes one value per dimension, thereby representing a distinct, potential MEV attack, which can then be instantiated on a specific DAG-based BFT protocol. We perform a set of experiments, each isolating a single dimension where the protocol permits it, to empirically measure its effect on the success rate of MEV attacks against six production DAG-based BFT protocols. Our experimental evaluation reveals that every protocol we evaluate is vulnerable to at least a subset of the MEV attacks in this space, and that which attacks succeed is mostly dictated by the protocol's own design rather than by attacker effort.
\end{abstract}

\section{Introduction}\label{sec:intro}

Byzantine fault-tolerant (BFT) consensus protocols are the core engines powering the state machine replication (SMR)~\cite{lamport1978time,schneider1990implementing} paradigm, ensuring that non-faulty replicas execute client requests in the same order (\emph{safety}) and every valid request is eventually executed (\emph{liveness}), despite the existence of up to $f$ Byzantine replicas.
In the classical design, one replica is chosen as the \emph{proposer}, which collects transactions from clients, packs them into an ordered batch, and
sends that batch to everyone else. A transaction's position in the batch is its execution order. This quietly hands the proposer the power to decide which transactions go in and in what sequence, and it can do both without breaking safety or liveness.

That power turns into profit when the application on top holds money~\cite{nakamoto2008bitcoin,wood2014ethereum}. In decentralized finance, a proposer can reorder trades for gain, a practice known as \emph{maximal extractable value} (MEV)~\cite{daian2020flash,eskandari2019sok,klages2019stability,qin2022quantifying,zhou2021high,baum2021sok,heimbach2022sok}. The standard example is the \emph{sandwich} attack: on seeing a victim's large trade against an automated market maker~\cite{xu2023sok}, the proposer inserts its own purchase just before the victim's trade and its sale just after, capturing the price movement the victim's own trade produced. Nothing is forged, no signature is broken, and no rule is violated; the attacker only exercises choices the protocol already grants it. Both legs depend on the ordering rule: getting in \emph{front} of the victim, and getting back in \emph{close behind} it before a competitor claims the same opportunity.
Ordering games of this kind have taken more than \$1.3 billion from ordinary users on Ethereum alone~\cite{mev2023chainlink}, and a measurement study of eleven million blocks found close to $200{,}000$ such attacks~\cite{torres2021frontrunner}. Nor is the practice confined to one chain or one layer: the same extraction has been measured across layer-2 rollups~\cite{torres2024rolling}, and the same composability that enables it also funds it, since an attacker can borrow the capital needed for an attack and repay it within the same transaction~\cite{qin2021attacking}. The practice has extended to newer chains as they have grown: on Sui, third-party infrastructure built specifically to mitigate the problem reports MEV extraction of $\sim\$18{,}000$ per day~\cite{shio}.

Traditional BFT protocols route all transaction dissemination through a single node, the proposer, which becomes a throughput bottleneck as the committee grows. \emph{DAG-based} BFT protocols~\cite{keidar2021all,danezis2022narwhal,spiegelman2022bullshark,giridharan2024autobahn,cheng2024shardag} remove that bottleneck by letting every validator propose blocks in parallel. Each block references earlier blocks; these references form a {\em directed acyclic graph (DAG)}, and the final transaction order is computed from that graph by a rule every validator runs identically. DAG-based protocols have seen widespread production use, with notable examples such as Sui \cite{sui} (carrying more than \$150 billion in cumulative on-chain exchange volume and powering more than $900$ applications~\cite{defillama_sui}), Aptos \cite{aptos}, Celo \cite{celo}, Chainlink~\cite{chainlink}, and Supra~\cite{supra}.
The flexibility inherent to DAG-based designs, however, is also what introduces new MEV vulnerabilities. Each validator independently controls which blocks it references, when it proposes, and what it includes in its own block. Every one of these choices is one an honest validator could legally make, yet each can be used to bias the final order, leading to more complex inter-block relationships that an adversary can exploit.

Existing studies measure only isolated pieces of MEV on DAG-based protocols~\cite{zhang2024no,mahe2025order}, reporting results that are hard to reconcile: no two studies share a setup, they target different attacks and protocols, grant the attacker different budgets, and, critically, define success differently, e.g., whether the attacker's block merely lands earlier in the final order or must land in the same round as the victim's. As with the broader MEV literature, attacks are reported individually, on different protocols, and under different metrics, making it unclear whether two attacks differ fundamentally or merely in how they are described. The field thus holds a collection of individually correct findings that do not combine into a comprehensive, comparable picture.

This paper closes that gap by presenting an attack space for MEV on DAG-based BFT protocols, organized around four families: the adversary (what the attacker does, with whom, and at what price), the protocol (how the target builds and linearizes its DAG), the target (who is attacked, how value is spread, and how success is measured), and the deployment (committee size, stake, and network latency). For each family, we identify the dimensions that shape an attack's impact. Each point in the attack space fixes one value per dimension, thereby representing a distinct, potential MEV attack, which can then be instantiated on a specific DAG-based protocol.

The full space is far too large to run exhaustively, and some parts of it are uninformative, so our contribution is not to enumerate it but to navigate it. We evaluate one representative point per dimension value, more than fifty in total, each varying a single dimension while holding the others fixed as far as the protocol allows. Each experiment instantiates its point on one or multiple candidate DAG-based protocols whose structures make that point feasible. In total, six production DAG-based BFT protocols (Narwhal-Tusk \cite{narwhalrepo2026}, Bullshark \cite{suirepo2026}, Mysticeti \cite{mysticetirepo2026}, AlephBFT \cite{alephbftrepo2026}, Mahi-Mahi \cite{mahimahirepo2026}, and Autobahn \cite{autobahnrepo2026}) appear across the experiments.

In summary, we introduce an attack space for MEV on DAG-based BFT protocols and evaluate a representative selection of points within this space across a range of candidate DAG-based protocols. Our experiments produce several noteworthy findings, a few of which we highlight below.

\begin{itemize}[leftmargin=1.2em,itemsep=3pt,topsep=3pt]
\item \textbf{Design decides exposure.} Every protocol we tested is vulnerable somewhere, and each vulnerability is highly dependent on architectural choices. Knowing how a protocol is built helps predict which attacks will succeed against it.
\item \textbf{Metrics change the verdict.} The definition of success significantly affects the measured success rate; for example, in a back-running attack, the rate depends heavily on how large a gap between the attacker's block and the victim's is still counted as a successful attack.
\item \textbf{Advantage without an attacker.} Some protocols grant an ordering advantage even when no attacker is present. For example, the Mysticeti ordering rule breaks ties within a round using the validator's identifier, so a low-numbered validator is systematically placed ahead of a high-numbered one.
\item \textbf{Budget substitutes for compromise.} A well-funded adversary does not necessarily need to compromise any validator to succeed; for example, a single malicious validator that bribes three honest validators achieves a higher success rate than four colluding validators. Acting alone, without bribing anyone, that same validator performs worse than not attacking at all. 
\item \textbf{Coordination is the multiplier.} Attackers who each target their own victim independently gain nothing from one another, so four of them score no better than one acting alone, whereas attackers who coordinate on a single victim gain with every member they add.
\end{itemize}
\section{System and Threat Model}\label{sec:model}

\noindent {\bf System model.}
A Byzantine Fault-Tolerant (BFT) protocol runs on a network of $3f+1$ replicas, at most $f$ of which may exhibit arbitrary, potentially malicious behavior. BFT protocols implement \emph{State Machine Replication} (SMR) \cite{lamport1978time,schneider1990implementing}, in which a service's state is replicated across a set of deterministic replicas. The goal of a BFT SMR protocol is to assign every request a position in the global order and to execute requests across all replicas in that order \cite{singh2008bft}.

This paper focuses on a class of BFT SMR protocols known as \emph{DAG-based BFT protocols}. Unlike traditional BFT protocols, DAG-based protocols decouple transaction dissemination from the consensus routine, splitting the process into two phases that run asynchronously: \emph{DAG construction} and \emph{total ordering}.
During DAG construction, replicas continuously disseminate transactions and organize them into a directed acyclic graph (DAG). Each block in the DAG contains a list of transactions together with references to blocks from the previous round, encoding its causal history. This phase proceeds independently of the total-ordering phase, and replicas keep extending the DAG without waiting for agreement to be reached.
During total ordering, a deterministic \emph{commit rule} designates a round leader. If a leader's block accumulates sufficient support from the DAG, it is committed together with its entire causal history, forming a \emph{committed sub-DAG}. Each committed sub-DAG is then \emph{linearized} into a total order of transactions to be executed by the execution layer.

As with BFT SMR protocols in general, a DAG-based BFT protocol must guarantee \emph{safety} (all non-faulty replicas execute the same requests in the same order) and \emph{liveness} (every request submitted by a correct client is eventually executed).
In an asynchronous system where replicas may fail, no deterministic consensus protocol can guarantee both safety and liveness (the FLP result)~\cite{fischer1985impossibility}. DAG-based BFT protocols circumvent this impossibility in one of two ways: by assuming a \emph{partial synchrony} model~\cite{spiegelman2022bullshark,babel2025mysticeti,shrivastav2019shoal,arun2025shoal++,shrestha2024sailfish}, or by introducing \emph{randomization}~\cite{keidar2021all,danezis2022narwhal,spiegelman2022bullshark,dai2023gradeddag,hu2026lemonshark}.
Under partial synchrony, the system is assumed to eventually stabilize: after an unknown global stabilization time (GST), messages between correct replicas are delivered within a known bound $\Delta$, and liveness is guaranteed only from that point on.
Under randomization, liveness instead holds probabilistically, independent of network timing. Each round, an unpredictable common coin elects the round leader only after that round's DAG has already been built, so a scheduling adversary cannot delay messages in anticipation of who will lead. With constant probability, the elected leader's block has already gathered enough DAG support to commit, so termination happens in an expected constant number of rounds, guaranteeing liveness with probability 1 even under full asynchrony.

Beyond the choice of synchrony model, DAG-based BFT protocols inherit the standard system assumptions of classical BFT protocols. Faulty clients, unlike faulty replicas, are unbounded in number. Replicas communicate over an unreliable, point-to-point network of bi-directional channels that may drop, corrupt, or delay messages. Finally, the adversary is strong: it may coordinate the behavior of malicious replicas and adaptively delay message delivery, but it cannot break the underlying cryptographic assumptions.

\noindent {\bf Threat model.}
BFT protocols are typically analyzed against a Byzantine adversary that may deviate arbitrarily from the protocol. Our focus, however, is on MEV: the value a validator can capture by positioning transactions in the committed order. We therefore adopt a \emph{rational} adversary that behaves strategically to maximize extracted value while remaining within protocol-legal behavior.

The adversary is confined to a \emph{legitimate action set} $\mathcal{L}$, the set of choices an honest validator could itself legally make under the protocol: (i) \emph{which} valid parent blocks it references when proposing a block, subject to the required $2f{+}1$ quorum; (ii) the \emph{content and internal ordering} of transactions within its own block; and (iii) \emph{when} it broadcasts its block, within the protocol's timing bounds. In this context, the adversary never violates consensus logic: it does not alter validity predicates or quorum-intersection checks, forge signatures, equivocate, or otherwise break safety. Denial-of-service attacks, liveness attacks, and safety violations lie entirely outside $\mathcal{L}$. Within $\mathcal{L}$, the adversary is otherwise unconstrained: colluding validators may coordinate freely, and any validator may exploit public information, including a protocol's leader schedule where that schedule is itself public and predictable (e.g., round-robin). Honest validators run unmodified binaries throughout the analysis.

Within $\mathcal{L}$, the adversary may also pay. A {\em bribe} is an offer to a validator that would otherwise follow its default behavior, in exchange for making one specific choice from $\mathcal{L}$ rather than another (omitting one victim's parent references from a proposal in our evaluation). Because the purchased action was permitted, accepting a bribe requires no deviation from consensus logic and the resulting messages are indistinguishable from those of a validator that simply referenced a different valid parent set. We do not model what a bribe costs, as pricing a validator's willingness to be paid would require assumptions about stake, reputation, and slashing outside this paper's scope. We instead measure what a bribe \emph{achieves}.

\ifextend Restricting the adversary to $\mathcal{L}$ isolates the contribution of protocol design from that of raw Byzantine power: every result in this paper demonstrates value extractable \emph{without breaking any protocol rule}, placing the responsibility for MEV extraction on the ordering mechanism rather than on an over-strong adversary model.\ \fi
\section{The MEV Attack Space}\label{sec:space}

\ifextend
Maximal Extractable Value (MEV) denotes the value a validator captures by controlling the position of transactions within the committed order. That control is not an abstraction. On Sui, two transactions that touch the same object run in the order the DAG committed them, so if both are trading against the same pool, the one placed earlier trades first and the later one gets the worse price. Position is how the money is made.
\ \fi

Order manipulation, and maximal extractable value (MEV) in particular, has been studied extensively
on leader-based chains, where a proposer front-runs, back-runs or sandwiches a victim for profit
\cite{eskandari2019sok,daian2020flash,klages2019stability,qin2022quantifying,zhou2021high,baum2021sok,heimbach2022sok}.
That line of work assumes one proposer per block, so the contest it studies is reordering inside a
single proposer's batch.
A second line proposes order-fairness or MEV-resistant mechanisms for classical BFT
\cite{zhang2020byzantine,kursawe2020wendy,kursawe2021wendy,kelkar2020order,kelkar2023themis,cachin2022quick,nagda2024rashnu},
and recent work carries the idea to DAG-based protocols \cite{nagda2026dag,kang2025fairdag}. These fix
an ordering property by construction rather than measuring what an attacker achieves against a rule
already deployed.
Closest to us, Zhang and Kate \cite{zhang2024no} identify DAG-specific strategies, fissure,
speculative and sluggish, on Narwhal-Tusk and Bullshark, and Mah\'e and Tucci-Piergiovanni
\cite{mahe2025order} show that a minority group can violate order fairness in a simulated DAG-Rider
by manipulating DAG construction and reliable broadcast. Both report a few attacks on a few
protocols under one definition of success. \ifextend Those strategies have since been benchmarked across seven implementations~\cite{mirzaei2027benchmark}, and the identifier-based ordering bias and gas-price re-sort of one uncertified design measured in detail~\cite{mirzaei2026fair}.\ \fi The space we set out below holds those strategies as
three values of a single dimension and adds the dimensions they keep fixed: which protocol property
an attack consumes, whom the attacker works with and at what price, and how success is defined,
which turns out to move the reported number furthest.
Our \emph{attack space} is spanned by four families: \textbf{A}dversary, \textbf{P}rotocol, \textbf{T}arget, and \textbf{D}eployment, each holding \emph{dimensions} along which an MEV attack can vary. Table~\ref{tab:taxonomy} lists them with their values. A point fixes one value per dimension and thereby represents a distinct, potential MEV attack, which can then be instantiated on a specific DAG-based BFT protocol.
\ifextend Changing one value turns one known attack into a related but distinct one. Along Relationship (A\ref{dim:relation}), for instance, a lone attacker becomes \emph{competing} once several act independently against different victims, \emph{colluding} once they agree out of band on one shared victim, and \emph{multi-group} once two such colluding groups, each with its own victim, compete for the same positions. Along Incentive (A\ref{dim:incentive}), that same colluding group gains \emph{bribery} once it pays otherwise-honest validators to join rather than controlling them outright. Each step changes exactly one dimension.\ \fi
Not every point represents a real attack. Some points are impossible because their values cannot hold at once: setting A\ref{dim:relation} to \emph{multi-group} while setting A\ref{dim:fraction} to $\alpha = 1/13$ asks for several groups of attackers in a network holding one attacker node\ifextend, and a point that asks for the \emph{censor} action to be scored by the \emph{all-pairs} metric asks where a block sits when the whole point of the action is that it never arrives\fi. A point can also be perfectly coherent yet impossible to instantiate on a given protocol, because that protocol lacks the component the point names: a point whose DAG type is \emph{certified} has nothing to run on in Mysticeti.

This section does not aim to enumerate every dimension; rather, it demonstrates the methodology used to define them.

\begin{table}[tbp]
\centering\footnotesize
\caption{The MEV attack space: families, dimensions and values.}
\label{tab:taxonomy}
\setlength{\tabcolsep}{1pt}%
\begin{tabular}{@{}p{0.03\columnwidth}>{\raggedright\arraybackslash}p{0.27\columnwidth}>{\raggedright\arraybackslash}p{0.65\columnwidth}@{}}
\toprule
& Dimension & Values \\
\midrule
\textbf{A} & A\ref{dim:action} Action & front-run, back-run, sandwich, censor \\
 & A\ref{dim:primitive} Primitive & fissure, speculative, sluggish, silent-except-leader, SLW, proposal-timestamp, LVW, vote-withhold \\
 & A\ref{dim:relation} Relationship & solo, competing, colluding, multi-group \\
 & A\ref{dim:info} Information & local, global, oracle \\
 & A\ref{dim:incentive} Incentive & none, bribery \\
 & A\ref{dim:fraction} Fraction & $\alpha$: $1/13$, $4/13$, $6/13$ (in our evaluation) \\
\midrule
\textbf{P} & P\ref{dim:type} DAG type & certified, uncertified \\
 & P\ref{dim:ordering} Ordering rule & leader-anchored, round-only, slot-based, election-based \\
 & P\ref{dim:tiebreak} Tiebreak & author, round, digest, seeded, order-fair \\
 & P\ref{dim:leader} Leader rule & rotation, stake, reputation \\
 & P\ref{dim:tuning} Tuning & protocol-specific and open-ended (e.g., batch size, max delay, wave length, GC depth, slot budget,
   election lookahead), each with its own values \\
\midrule
\textbf{T} & T\ref{dim:victim} Victims & single, multiple \\
 & T\ref{dim:metric} Metric & all-pairs, same-round, committing-height, realized MEV,
        $L_1$, $L_2$, triplet sandwich, inclusion \\
 & T\ref{dim:value} Value & uniform, pareto, lognormal \\
\midrule
\textbf{D} & D\ref{dim:size} Size $n$ & $13$, $25$, $49$ (in our evaluation) \\
 & D\ref{dim:stake} Stake & equal, skewed \\
 & D\ref{dim:geo} Geo-distribution & LAN, WAN, asymmetric \\
\bottomrule
\end{tabular}
\end{table}

\subsection{The Adversary Family}\label{sec:space:adversary}
Family A is the heart of the space, so we describe it first and in the greatest detail.
\begin{adver}
\label{dim:action}
\textbf{Action}. Action describes the position the attacker wants relative to a victim block, or, for censorship, the absence of one.
{\em Front-running} is the attempt to place an attacker block before a victim block so that the attacker's transactions execute first.
{\em Back-running} targets the opposite side of the victim block in order to trade on the state the victim's transaction just created.
A {\em sandwich} attack combines front-running and back-running around the same victim block.
{\em Censorship} removes a transaction rather than repositioning it.
\end{adver}

\begin{figure*}[tbp]
  \centering
  \begin{minipage}[t]{0.32\textwidth}
    \centering
    \includegraphics[width=0.94\textwidth]{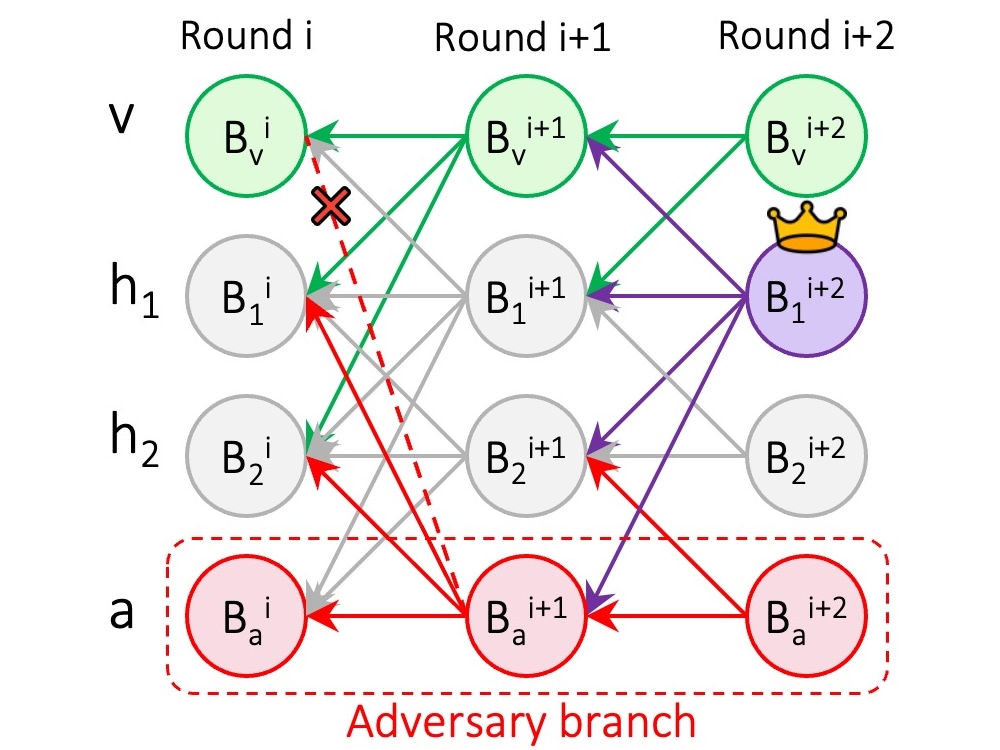}\\[2pt]
    \small (a) Fissure
  \end{minipage}\hfill
  \begin{minipage}[t]{0.32\textwidth}
    \centering
    \includegraphics[width=0.94\textwidth]{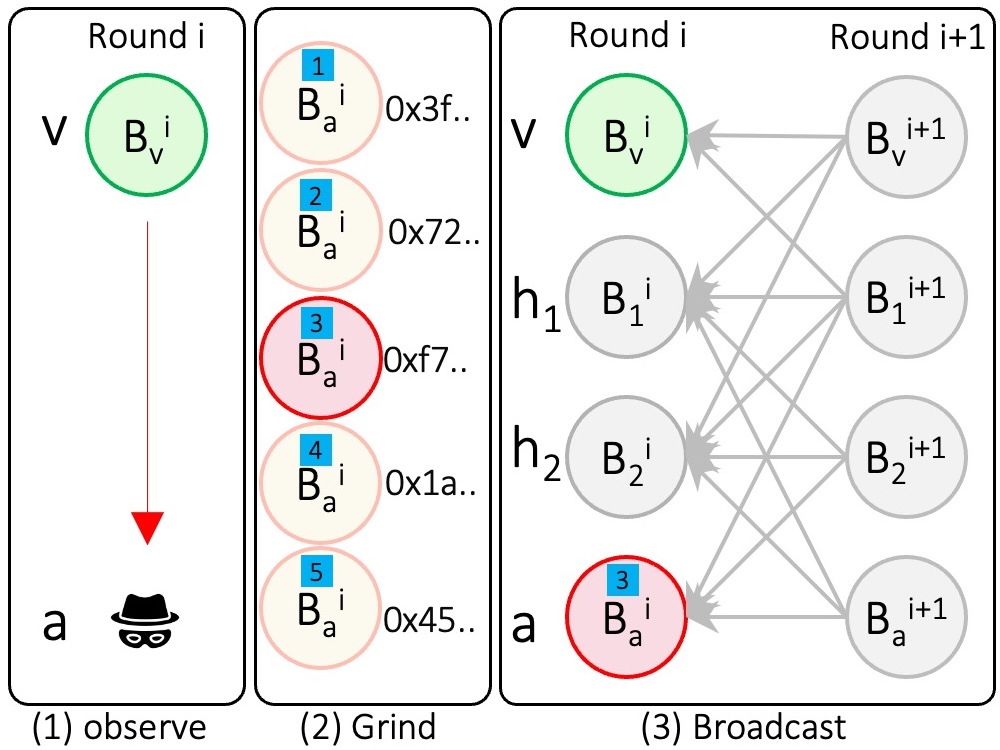}\\[2pt]
    \small (b) Speculative
  \end{minipage}\hfill
  \begin{minipage}[t]{0.32\textwidth}
    \centering
    \includegraphics[width=0.94\textwidth]{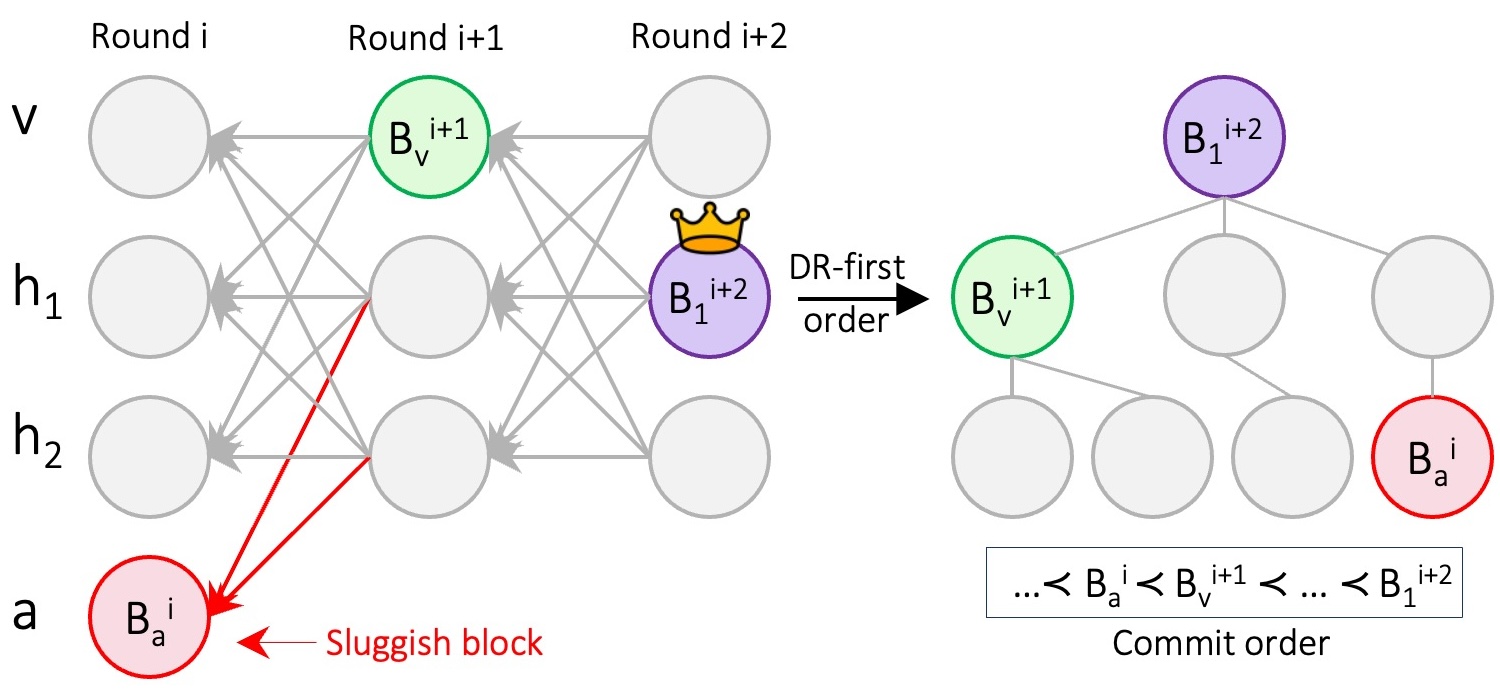}\\[2pt]
    \small (c) Sluggish
  \end{minipage}
  \caption{One representative move from each lever. In (a), the attacker refuses to reference the victim's block, so the victim gathers less support and is committed later. In (b), the attacker privately builds several valid candidate blocks and broadcasts only the one that sorts best. In (c), the attacker delays its own block so that the round-first ordering rule still places it ahead of the victim.}
  \label{fig:mechanics}
\end{figure*}

\begin{adver}
\label{dim:primitive}
\textbf{Primitive}.
The primitive is the specific move a validator makes. We group primitives by the \emph{lever} they pull, which lets us anticipate which protocols a given primitive can affect. Figure~\ref{fig:mechanics} illustrates one representative move from each lever.

\begin{itemize}[leftmargin=1em,itemsep=1pt,topsep=0pt]
\item \uline{Attacks on another validator's position}, which withhold the support a victim's block needs. The protocols count that support at more than one moment, and which moment a primitive targets decides where it works. \emph{Fissure} omits a victim's block from the attacker's parent set (Fig.~\ref{fig:mechanics}a). It targets the quorum that \emph{admits} a block, so it works where a block must gather a certificate before it counts and does nothing where blocks are admitted on arrival. \emph{Leader-vote withholding} (LVW) instead withholds a vote that a \emph{leader} needs in order to commit: the attacker proposes as usual, but leaves the leader's block out of its own parent set, so its block counts as a refusal rather than as support. On the uncertified design that vote is simply a parent reference in the following round, which is why the move exists there at all; on a certified design the same refusal is fissure again rather than a separate primitive. \emph{Vote-withholding} acts one step earlier than either: on a certified design the attacker declines to sign the victim's header at all, so no certificate ever forms. Fissure moves a victim later in the order; withholding the signature can keep it out of the order altogether, which is why censorship is measured on this move rather than on fissure.

\item \uline{Attacks on the attacker's own position}, which change \emph{when} the attacker's blocks appear. These attacks succeed only when the protocol makes some rounds positionally more valuable than others. A leader-anchored linearization has this property, since a leader's block heads the sub-DAG it commits; a linearization that orders batches by round alone does not, and the attack yields no advantage. \emph{Sluggish} keeps every slot but delays its proposals into more favorable ones (Fig.~\ref{fig:mechanics}c). \emph{Silent-except-leader} speaks \emph{only} in the rounds the attacker leads and stays quiet in every other round, so every block it publishes lands at a commit anchor. \emph{Strategic leader withholding} (SLW) does the reverse: the attacker speaks in every round \emph{except} the one it leads. Skipping its own slot forces the other validators to wait out the leader timeout, so that wave is folded into the next leader's history and the boundary between two commits moves.

\item \uline{Attacks on a block's contents}, which change what the attacker's own block holds while staying within protocol rules. In particular, \emph{Speculative} privately generates several valid candidate blocks and publishes only the one yielding the best ordering (Fig.~\ref{fig:mechanics}b), while \emph{proposal-timestamp} stamps the block with a legal but advantageous timestamp.
\end{itemize}

\ifextend\noindent Silent-except-leader and SLW both govern \emph{when the attacker itself speaks}, but in opposite directions: one speaks only when it leads, the other only when it does not. LVW is not about the attacker's own turn at all; it is about refusing to support \emph{another} validator's leader block.\ \fi

\end{adver}

\begin{adver}
\label{dim:relation}
\textbf{Relationship.} Relationship is how the attacking validators stand toward one another: whether they agree on a plan, and whether they aim at the same victim.
\emph{Solo} is when there is only one attacking validator.
\ifextend Attacks that reposition the attacker's own blocks are already fully effective here.\ \fi
\emph{Competing} is when several attackers act independently, each choosing its own victim, with no agreement between them.
\ifextend Their efforts do not add up: spread across separate victims, each attacker's denial looks like ordinary noise.\ \fi
\emph{Colluding} is when several attackers act as one group against one victim, agreeing out of band on the target.
\ifextend This is the only arrangement that lets denial accumulate.\ \fi
Finally, \emph{Multi-group} is when two or more colluding groups, each with its own victim, compete for the same positions.
\ifextend Each group pays the coordination cost while facing rivals for the advantage it is trying to buy.\fi
\end{adver}

\begin{adver}
\label{dim:info}
\textbf{Information.} Information is what the attacker must \emph{observe} before its primitive works.
\ifextend The dimension is graded by where the knowledge comes from, not by how secret it is: every value below is obtainable without privileged access, and they differ in how much of the system a validator has to watch to obtain it.\ \fi
\emph{Local}: only the attacker's own view, meaning the blocks it has received and its own state.
\ifextend Most primitives need nothing more, which is why they work at all.\ \fi
\emph{Global}: committee-wide facts that are public but must be assembled, such as the protocol's configured parameters together with what the current round actually contains.
\ifextend Cheap to obtain, and enough on its own to decide whether a mitigation holds.\ \fi
\emph{Oracle}: information the protocol never carries, supplied from outside it, such as the value of a pending transaction.
\ifextend This value exists because chains without a public mempool (Sui, Aptos) give an attacker no in-protocol way to see what a pending transaction is worth, so there the \emph{Speculative} primitive must be oracle-fed rather than mempool-fed.\fi
\end{adver}

\begin{adver}
\label{dim:incentive}
\textbf{Incentive.} Incentive is whether the attack relies only on the validators the attacker already controls (\emph{none}), or additionally pays an \emph{honest} validator to take one specific, protocol-permitted action on its behalf (\emph{bribery}), e.g., omitting one victim's parent reference from one proposal.
\ifextend The attacker does not buy the validator's keys, does not make it run attacker code, and does not ask it to break a rule. In particular, omitting references is the same move \emph{Fissure} makes and is a choice any proposer may lawfully make. The consequence is that the protocol cannot tell the two apart: a bribed validator emits exactly the messages of an honest validator that happened to reference a different parent set, so nothing in the DAG records that a payment occurred. A bribe therefore buys \emph{participants}, and participants only help an attack that needs a coalition. \fi
\end{adver}

\begin{adver}
\label{dim:fraction}
\textbf{Fraction.}
The fraction, denoted $\alpha$, represents the proportion of validators controlled by the attacker. \ifextend In contrast to A\ref{dim:relation}, this dimension varies the fraction of attackers while keeping the relationship fixed.\ \fi
In our evaluation, we report $\alpha \in \{1/13,\,4/13,\,6/13\}$, with $\alpha = 4/13 = f/n$, as the default operating point.
\end{adver}

\subsection{The Protocol Family}\label{sec:space:protocol}

Most dimensions in the P family are decisions the protocol's designer makes once, at design time, and an operator cannot change them afterward; the one exception, Tuning (P\ref{dim:tuning}), is exactly the kind of choice an operator does control.

\begin{proto}
\label{dim:type}
\textbf{DAG type.}
DAG type determines whether a block must collect a quorum of endorsements before it can enter the DAG at all. In a \emph{certified} DAG, a block must gather a threshold of validator signatures before it counts, and it is this certificate (not the raw block) that other validators reference. What matters here is that some quorum must sign before the block is referable, not the size of that quorum: the usual threshold is $2f{+}1$, but Autobahn admits a block on a proof of availability of $f{+}1$ votes. In an \emph{uncertified} DAG, blocks are referenced directly on arrival, reducing the latency but permitting more transient disagreement about which blocks exist. The distinction matters for MEV because certification is an \emph{admission gate} that can be leveraged by an attacker. Withholding endorsement from a victim's block denies it something the victim needs in order to be seen at all. The identical refusal against an uncertified design denies nothing, since the block was already admitted on its arrival.

\ifextend This admission gate is worth separating from a second quorum that both constructions share, since the two are easy to conflate and different attacks target each one. The \emph{admission quorum} decides whether a block enters the graph at all; only a certified DAG has one. The \emph{commit quorum} comes later and decides when a leader's history is emitted, a leader commits only once enough subsequent blocks reference it, and every construction here requires it, certified or not. A refusal to reference a victim therefore means two different things: against admission, it can exclude a block outright; against a commit, it can only delay a leader, and only if enough validators refuse together.\fi
\end{proto}

\begin{proto}
\label{dim:ordering}
\textbf{Ordering rule.} The ordering rule totally orders the committed sub-DAG into a sequence. \emph{Leader-anchored} schemes place the leader's own block first, followed by everything it referenced. \emph{Round-only} schemes place blocks strictly by round number, so no block in a round is placed ahead of the others in that same round. \emph{Slot-based} schemes build the sequence slot by slot, taking one block from each validator's own chain in turn, leaving a single proposer little room to move. \emph{Election-based} schemes instead run a separate election to pick which block leads, shifting the contest from proposing to winning that election.
\end{proto}

\begin{proto}
\label{dim:tiebreak}
\textbf{Tiebreak.} The ordering rule leaves ties unresolved: blocks from the same round must be separated somehow, and each protocol fixes its own rule for doing so. \emph{Author-based} schemes sort by a fixed, pre-assigned validator identity (author index). \emph{Round-based} schemes sort on the round number alone and leave blocks of the same round in whatever order the traversal yields, so the rule carries no identity of its own. \emph{Digest-based} schemes sort by the hash of each block. \ifextend This no longer favors a fixed identity, but a proposer can still search over it, since it controls its own block's contents.\ \fi \emph{Seed-based} schemes combine the hash with a value drawn per commit, so the sort key is fixed only after the block is built. \emph{Order-fair} schemes use the order in which validators received the blocks \cite{nagda2026dag,kang2025fairdag}.
\end{proto}

\begin{proto}
\label{dim:leader}
\textbf{Leader rule.} The leader rule decides which validator leads a given round. It matters wherever the ordering rule favors leaders (e.g., leader-anchored), since it determines how often, and how predictably, an attacker occupies that position. \emph{Rotation} schemes pass leadership from validator to validator. The rotation may be fixed, as in a round-robin cycle, in which case every validator knows in advance which rounds are its own; or random, as when a shared coin draws the leader, in which case the schedule cannot be anticipated. \emph{Stake-weighted} schemes assign leadership in proportion to stake, so leader positions can effectively be bought. \emph{Reputation-based} schemes determine the leader from recent behavior, making the schedule depend on something an attacker cannot set directly.
\end{proto}

\begin{proto}
\label{dim:tuning}
\textbf{Tuning.} P1--P4 are structural: a designer fixes them once, and an operator cannot change them afterward. Every deployed protocol also exposes \emph{tuning parameters} that an operator sets at run time, and these turn out to move MEV exposure as much as some structural choices do. Because they are protocol-specific, we group them into a single, open-ended dimension. Some examples include worker batch size and maximum batching delay (e.g., Narwhal-Tusk), wave length in the wave-based protocols (e.g., Mahi-Mahi), garbage-collection depth, cache depth, synchronization timeout, and election lookahead (e.g., AlephBFT), and the concurrent slot budget (e.g., Autobahn). This list is not exhaustive, and it is not meant to be: the point of the dimension is that every protocol carries parameters of its own that can shift an attack's success rate.
\end{proto}

\subsection{The Target Family }\label{sec:space:target}

The target family is what the attack aims at, and how we decide whether it worked.

\begin{target}
\label{dim:victim}
\textbf{Victims.} How many validators the attacker targets at once. \emph{Single} focuses the attacker's entire effort on a single victim. \emph{Multiple} targets several victims simultaneously, which splits that effort: the attacker still needs enough parent references to meet quorum, so excluding more victims leaves fewer non-victim validators to draw on instead.
\end{target}

\begin{target}
\label{dim:metric}
\textbf{Metric.} Metric defines whether a placement counts as a success. A committed order is a sequence of blocks, each block $b$ with a proposer $\pi(b)$, a position $p(b)$ in that sequence, and a round $r(b)$. Let $A$ be the attacking validators and $V$ the victims, and write $B_A$ and $B_V$ for the blocks each proposes. We start with metrics that can be used for front-running attacks (all-pairs, same-round, targeted committing-height, and realized MEV) and continue with back-running ($L_1$, $L_2$), sandwiching (triplet sandwich), and censorship (inclusion).

\emph{All-pairs (the positional metric)}: the fraction of attacker--victim block pairs in which the attacker's block is placed before the victim's, over all such pair in the committed order.
\[
\textsf{ASR}_{\mathrm{ap}} =
\frac{\bigl|\{(b_a,b_v)\in B_A\times B_V : p(b_a) \prec p(b_v)\}\bigr|}{|B_A|\cdot|B_V|}
\]
Under a fair (order-unbiased) linearization, each pair is equally likely to fall either way, so $\textsf{ASR}_{\mathrm{ap}}{=}50\%$ is the exact neutral baseline. We take all-pairs ASR as the paper's default \emph{positional} metric.

\emph{Same-round (the classical front-run rate)}: how often a victim block $b_v$ has an attacker block $b_a$ ahead of it in the same round ($r(b_a) = r(b_v)$). The same-round attack success rate is the fraction of victim blocks that are front-run.
\[
\textsf{ASR}_{\mathrm{sr}}=
\frac{\bigl|\{b_v {\in} B_V: \exists b_a \in B_A, p(b_a) {\prec} p(b_v), r(b_a) = r(b_v)\}\bigr|}{|B_V|}
\]
This is the standard MEV predicate (an attacker can only sandwich or front-run a victim it is committed \emph{alongside}), but it is \emph{not} a neutral baseline, e.g., on a protocol that breaks ties inside a round by author index, a low-numbered attacker front-runs nearly every same-round victim by design. \ifextend This is the clearest illustration of why a metric cannot be interpreted without knowing the protocol it was measured on.\ \fi

\emph{Targeted committing-height}: this metric scores only the attacker--victim block pairs in which the attacker block specifically targeted that victim block~\cite{zhang2024no}. For each such targeted pair $(b_a, b_v)$, the attack succeeds when $p(b_a) < p(b_v)$, and the rate is taken over targeted pairs where both blocks committed.

\emph{Realized MEV}: Rather than counting victim blocks equally, this metric weights each one by value, so a successful front-run of a high-value block counts for more than one of a low-value block. Let $w(b_v)$ be the value of victim block $b_v \in B_V$, and let $\mathbf{1}[b_v \text{front-run}]$ be $1$ if $b_v$ was successfully front-run under the same-round metric, and $0$ otherwise. Then
\[
\textsf{ASR}_{w} = \frac{\sum_{b_v \in B_V} w(b_v)\cdot \mathbf{1} [b_v \text{ front-run}]}{\sum_{b_v \in B_V} w(b_v)}.
\]
The denominator is the total value held by every victim block in $B_V$; the numerator restricts that same sum to the victim blocks that were successfully front-run. $\textsf{ASR}_w$ is therefore the fraction of victim value the attacker actually captured.
\ifextend Same-round is the right predicate because front-running requires a shared commit: to front-run a victim's transaction, the attacker's transaction must land in that same commit, just ahead of it. A block committed earlier in some other round is nowhere near the victim, so there is nothing to trade against.\ \fi

\emph{$L_1$ and $L_2$ (the back-running metrics).} A gap between the victim's block and the attacker's is dangerous to the attacker, since any block landing between them can seize the same state-dependent opportunity first~\cite{daian2020flash,zhou2021high}. We therefore define $L_1$, a successful back-run, as the fraction of victim blocks {\em immediately} followed by an attacker block with nothing in between, and $L_2$ as the fraction followed by at most one intervening block. Writing $L_k$ for the rate that allows at most $k-1$ blocks between the two:
\[
\textsf{ASR}_{L_k} = \frac{\bigl|\{b_v \in B_V \;:\; \exists\, b_a \in B_A,\; p(b_a) - p(b_v) \in \{1,\dots,k\}\}\bigr|}{|B_V|}
\]
The denominator counts victim blocks rather than attacker-victim pairs, so back-running is scored per victim.

\emph{Triplet sandwich (the sandwiching metric).} Sandwiching needs both sides of the bracket around the \emph{same} victim, so it is scored over comparable triplets of a front-attacker block $b^f_a$, a victim $b_v$, and a back-attacker block $b^b_a$:
\[
\textsf{ASR}_{\mathrm{sw}} = \frac{\bigl|\{(b^f_a, b_v, b^b_a) \;:\; b^f_a \prec b_v \prec b^b_a\}\bigr|}{\bigl|\{\text{comparable } (b^f_a, b_v, b^b_a)\}\bigr|}.
\]
 The denominator counts comparable triplets rather than victim blocks, so a rate near $100\%$ means the attacker closes nearly every bracket available to it, not that every victim was sandwiched.

\emph{Inclusion (the censorship metric).} No positional metric can score censorship: a block that never enters the committed order has no position to measure. We therefore score it on \emph{inclusion}, counting blocks instead of positions. Let $C(B_V)$ be the number of victim blocks committed, and $\tilde{C}(B_H)$ the median of that same count over the honest, non-victim validators. The censorship rate measures how far short of that median the victim falls:
\[
\textsf{ASR}_{\mathrm{cr}} = 1 - \frac{C(B_V)}{\tilde{C}(B_H)}.
\]
$\textsf{ASR}_{\mathrm{cr}} = 0$ means the victim was committed as often as everyone else and nothing was censored, while $\textsf{ASR}_{\mathrm{cr}} = 1$ means none of its blocks were committed at all.
\ifextend Values slightly below zero simply reflect the victim being committed a bit more often than the median, which is ordinary run-to-run variation rather than a signal. We use the median rather than the mean over honest validators because a run in which the attackers stall the protocol lowers every validator's count at once; a mean would let that stall read as censorship.\fi
\end{target}

\begin{target}
\label{dim:value}
\textbf{Value.} How much money each victim transaction is assumed to carry. This changes nothing about the order, so it affects only realized MEV discussed in T\ref{dim:metric}. \emph{Uniform} assigns every transaction the same value, so value tracks position exactly. \emph{Pareto} assigns most of the value to a few transactions, with the rest worth comparatively little. \emph{Lognormal} produces a similarly skewed distribution, so value again concentrates in a small number of transactions, though with a different tail shape than Pareto.
\end{target}

\subsection{The Deployment Family }\label{sec:space:deployment}
Deployment defines the operating environment. Among the many dimensions it could include, we consider these three: committee size, stake skew, and geo-distribution.

\begin{deploy}
\label{dim:size}
\textbf{Committee size.} The number of validators $n$ in the committee, which fixes the fault bound $f$ and, with it, each validator's share of the system. We run $n \in \{13, 25, 49\}$: $13$ is the smallest committee with $f{=}4$ and serves as our default, while $25$ and $49$ test whether an effect survives dilution. Size matters because leader anchors are shared out roughly as $1/n$, so a fixed attacker set commands a shrinking fraction of the valuable positions as the committee grows.
\end{deploy}

\begin{deploy}
\label{dim:stake}
\textbf{Stake skew.} How voting weight is distributed across the committee. \emph{Equal} assigns the same weight to every validator while \emph{skewed} concentrates the weight, giving one validator five times the average. This dimension asks whether buying weight buys ordering position, which depends on whether the leader rule (P\ref{dim:leader}) consults stake at all.
\end{deploy}

\begin{deploy}
\label{dim:geo}
\textbf{Geo-distribution.} Where validators sit relative to one another, and thus how long messages take between them. \emph{LAN} co-locates the committee; \emph{WAN} spreads it across regions with realistic inter-region delays; \emph{asymmetric} places one part of the committee far from the rest, so delay is distributed unevenly.
\ifextend The dimension matters because latency changes \emph{when} blocks arrive while the linearization rule stays deterministic, so only attacks whose lever is timing are sensitive to it at all. \fi
\end{deploy}
\section{Methodology}\label{sec:method}

\begin{figure*}[tbp]
\centering
\resizebox{\textwidth}{!}{
\begin{tikzpicture}[
  font=\footnotesize,
  box/.style   ={draw=black!30, rounded corners=3pt, align=left, inner sep=5pt},
  chip/.style  ={rounded corners=2pt, inner sep=3pt, font=\scriptsize, align=left},
  stage/.style ={font=\footnotesize\bfseries, align=right, anchor=east, text=black!75},
  badge/.style ={circle, draw=none, text=white, font=\bfseries\scriptsize, minimum size=5.4mm, inner sep=0pt},
  flow/.style  ={-latex, line width=1.8pt, draw=black!35},
  side/.style  ={-latex, line width=0.9pt, draw=black!35, densely dashed},
  hdr/.style   ={font=\footnotesize\bfseries}]

\node[badge, fill=cbSky!90!black] at (-4.35,-0.75) {1};
\begin{scope}[shift={(-3.55,-0.75)}]                     
  \foreach \i in {0,1,2}{\draw[fill=cbSky!35, draw=cbSky!70!black, line width=0.5pt]
    (-0.3,{-0.24*\i}) rectangle (0.3,{-0.24*\i+0.17});
    \fill[cbSky!80!black] (-0.22,{-0.24*\i+0.085}) circle (0.028);}
\end{scope}
\node[stage] at (-0.35,-0.75) {Setup \&\\Environment};
\node[box, fill=cbSky!15, text width=7.4cm, anchor=north west] (codebases) at (0,0)
  {\textbf{Instrumented Rust codebases.} Six deployed DAG-BFT protocols;\\
   \textbf{Bullshark} (certified) and \textbf{Mysticeti} (uncertified) measured first-hand};
\node[box, fill=black!4, text width=7.4cm, anchor=north west] (env) at (8.2,0)
  {\textbf{Run configuration.} Resource-capped containers on CloudLab;\\
   default $n{=}13$ ($f{=}4$), equal stake, 5 reps per cell, value model as an input};
\draw[side] (codebases.east) -- (env.west);

\node[badge, fill=black!55] at (-4.35,-3.15) {2};
\begin{scope}[shift={(-3.55,-3.15)}]                     
  \draw[fill=black!8, draw=black!45, line width=0.6pt]
    (0,0.42) -- (0.32,0.27) -- (0.32,-0.08)
    .. controls (0.32,-0.30) and (0.14,-0.40) .. (0,-0.46)
    .. controls (-0.14,-0.40) and (-0.32,-0.30) .. (-0.32,-0.08)
    -- (-0.32,0.27) -- cycle;
  \draw[draw=black!60, line width=0.9pt, line cap=round] (-0.13,0.03) -- (-0.03,-0.10) -- (0.16,0.17);
\end{scope}
\node[stage] at (-0.35,-3.15) {Attacker\\Injection \&\\Safety};
\node[box, fill=black!4, text width=15.6cm, minimum height=2.15cm, anchor=north west] (hooks) at (0,-2.15) {};
\node[anchor=north west] at (0.2,-2.24)
  {\textbf{Environment-gated hooks}, one per attacker behaviour, in three states:};
\node[chip, fill=cbGreen!22, anchor=north west, text width=4.5cm] at (0.25,-2.78)
  {\textbf{disarmed} $\rightarrow$ honest validator\\binary byte-identical to upstream};
\node[chip, fill=cbVermillion!22, anchor=north west, text width=4.8cm] at (5.15,-2.78)
  {\textbf{armed} $\rightarrow$ attacker\\choices confined to the action set $\mathcal{L}$};
\node[chip, fill=cbYellow!45, anchor=north west, text width=5.0cm] at (10.35,-2.78)
  {\textbf{armed + paid} $\rightarrow$ corrupted validator\\one lawful choice bought, not a Byzantine fault};
\node[anchor=north west, font=\scriptsize] at (0.25,-3.72)
  {\textbf{CI guardrail} rejects any change to consensus, validity, quorum or signature code};
\draw[flow] (codebases.south) -- (codebases.south |- hooks.north);

\node[badge, fill=cbSky!90!black] at (-4.35,-5.85) {3};
\begin{scope}[shift={(-3.55,-5.85)}]                     
  \draw[fill=cbSky!35, draw=cbSky!70!black, line width=0.5pt] (-0.32,-0.26) -- (-0.32,0.26) -- (0.0,0) -- cycle;
  \foreach \i in {0,1,2}{\draw[draw=black!45, line width=0.7pt, line cap=round]
    (0.12,{0.17-0.17*\i}) -- ({0.34-0.05*\i},{0.17-0.17*\i});}
\end{scope}
\node[stage] at (-0.35,-5.85) {Execution \&\\Logging};
\node[box, fill=cbSky!15, text width=6.8cm, anchor=north west] (run) at (0,-5.05)
  {\textbf{Run and record.} $n$ validator instances over the real network stack; the committed
   order of every run is logged};
\node[box, fill=cbVermillion!13, text width=8.2cm, anchor=north west] (verify) at (7.4,-5.05)
  {\textbf{Verification filter.} Every attack run must emit a marker proving its mechanism
   fired; every control is verified disarmed.\\
   \emph{Caveat:} the marker is only visible at raised log verbosity, and verbosity perturbs
   timing, so an attack and its control always share one setting};
\draw[flow] (run.north |- hooks.south) -- (run.north);
\draw[side] (run.east) -- (verify.west);

\node[badge, fill=cbGreen!75!black] at (-4.35,-8.85) {4};
\begin{scope}[shift={(-3.55,-8.85)}]                     
  \foreach \i/\h in {0/0.22,1/0.38,2/0.54}{\draw[fill=cbGreen!35, draw=cbGreen!70!black, line width=0.5pt]
    ({-0.32+0.23*\i},-0.28) rectangle ({-0.15+0.23*\i},{-0.28+\h});}
  \draw[draw=black!45, line width=0.6pt] (-0.38,-0.28) -- (0.34,-0.28);
\end{scope}
\node[stage] at (-0.35,-8.85) {Analysis \&\\Metrics};
\node[box, fill=cbGreen!13, text width=4.6cm, anchor=north west] (lift) at (0,-7.65)
  {\textbf{Lift over a paired control} taken at identical attacker and victim placement, one
   dimension varied at a time};
\draw[flow] (lift.north |- run.south) -- (lift.north);

\node[box, fill=black!3, text width=10.4cm, minimum height=3.0cm, anchor=north west] (band) at (5.2,-7.65) {};
\node[hdr, anchor=north, text=black!75] at (10.4,-7.82) {Metrics};
\node[chip, fill=cbBlue!16,       text width=2.6cm, minimum height=1.75cm, anchor=north west] at (5.45,-8.35)
  {\textbf{Positional}\\ same-round ASR\\ all-pairs ASR\\ targeted committing-height ASR};
\node[chip, fill=cbGreen!16,      text width=1.5cm, minimum height=1.75cm, anchor=north west] at (8.35,-8.35)
  {\textbf{Value}\\ realized MEV};
\node[chip, fill=cbVermillion!16, text width=2.4cm, minimum height=1.75cm, anchor=north west] at (10.1,-8.35)
  {\textbf{Post-victim}\\ $L_1$, $L_2$\\ triplet sandwich};
\node[chip, fill=cbYellow!35,     text width=2.4cm, minimum height=1.75cm, anchor=north west] at (12.7,-8.35)
  {\textbf{System}\\ inclusion};
\draw[side] (lift.east) -- (band.west);
\end{tikzpicture}}
\caption{How a row is measured}
\label{fig:method}
\end{figure*}
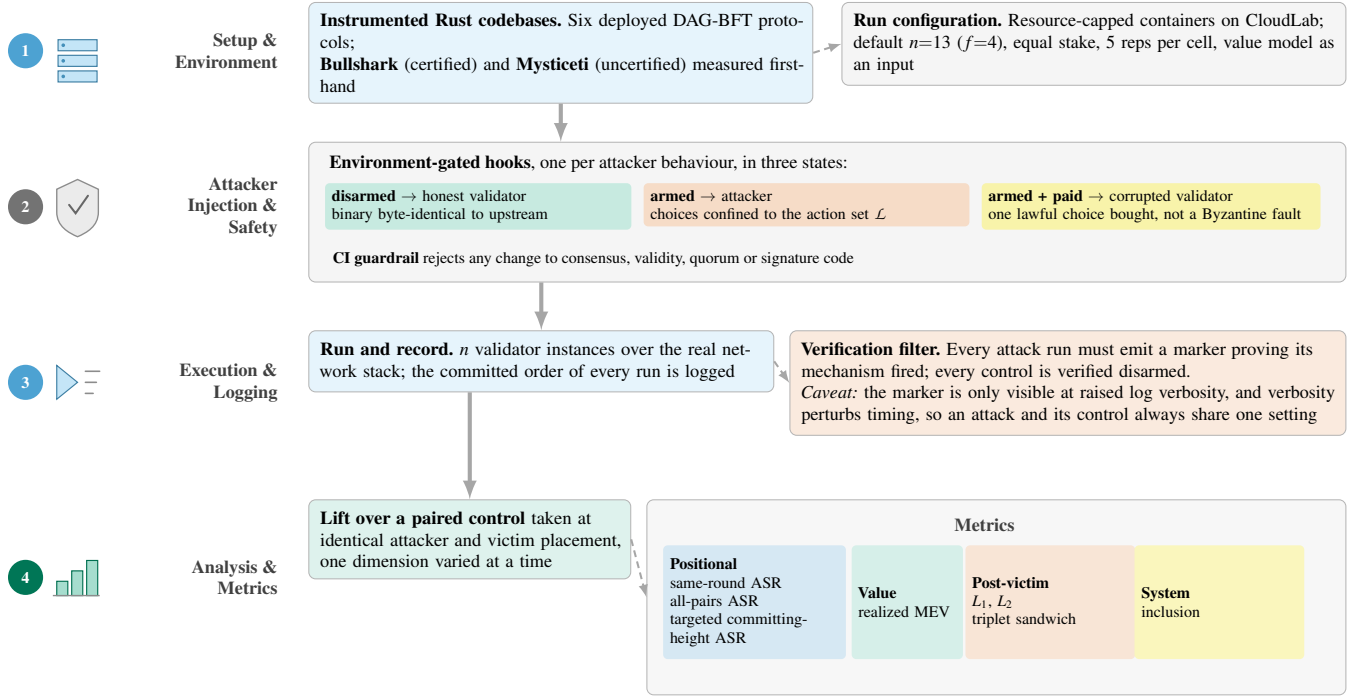

\subsection{Implementation and execution}
Figure~\ref{fig:method} summarizes how a single row is produced and scored. We instrument the production Rust implementations directly. Six DAG-BFT protocols appear across the experiments: Narwhal-Tusk, Bullshark, Mysticeti, AlephBFT, Mahi-Mahi, and Autobahn. A certified DAG (Bullshark) and an uncertified one (Mysticeti) carry most experiments. \emph{Bullshark}~\cite{spiegelman2022bullshark} runs its consensus over a separate
Narwhal~\cite{danezis2022narwhal} mempool: workers batch transactions, and a block enters the DAG only once $2f{+}1$ validators have signed it, so what a later block references is a certificate rather than the block itself. \emph{Mysticeti}~\cite{babel2025mysticeti}, whose reference design powers Sui in production, removes that step and references blocks directly, which is where its latency advantage comes from. These are the two designs that instantiate P\ref{dim:type}.

Two of the six are measured on \emph{two independent codebases}. Bullshark and Mysticeti each have a separate reference implementation and a deployment inside a larger production system, and we ran both. This is deliberate: agreement between two independent implementations of the same protocol separates a property of the protocol from an artifact of one codebase, and the two implementations do agree, within $1.1$ and $2.2$ points respectively on the primitive they share. Every attacker behavior in Family A is realized as an \emph{environment-gated hook} on the proposing path: when disarmed, the compiled binary is byte-identical to upstream, and honest validators always run the disarmed binary. A continuous-integration guardrail keeps every attacker inside the legitimate action set, meaning it may only make choices an honest validator could lawfully make, such as which blocks to reference, when to propose, and what to put in its own block. The guardrail rejects any change that touches consensus, validity, quorum, or signature code outside an
explicit attacker-hook allowlist, so no result can silently depend on breaking a protocol rule.

\paragraph{Evidence that a mechanism engaged.}
An attack hook that fails silently yields a plausible null result, and a control that silently still runs the attack yields a plausible zero lift. Both happened to us during this study. We therefore require every attack cell to emit a marker proving its mechanism fired; our scorer refuses to report a cell whose marker count is zero; and every control is verified to have actually disarmed the hook. We recommend the same discipline to anyone running this kind of experiment: the failure modes here produce publishable-looking numbers rather than obvious errors.

\subsection{Evaluation setting}
Each cell is a real multi-node consensus run: $n$ validator instances communicate over the actual network stack and reach consensus, and we record the committed order of every run. Nothing is simulated, and no order is reconstructed after the fact.
Unless stated otherwise, we use $n{=}13$ (the smallest committee with $f{=}4$) and equal stake. Each cell is repeated $5$ times, and we report the median. The full default configuration a cell varies one value away from is given in \S\ref{sec:find}. Runs execute in resource-capped containers on our dedicated CloudLab nodes, and every arm of a comparison runs on the same machine under the same caps, so a difference between two arms cannot come from the hardware they happened to land on.

Two things are held fixed between an attack run and its control, and they are what make the difference between them readable. The same validators are labeled attacker and victim in both, so any advantage those positions carry is present on both sides and cancels. And only one dimension moves at a time: committee size, stake skew, and wide-area latency are each injected on their own rather than together.
\setcounter{adver}{0}
\setcounter{proto}{0}
\setcounter{target}{0}
\setcounter{deploy}{0}

\section{Experimental Study}\label{sec:find}

\definecolor{Headline}{HTML}{F4F4F4}
\begin{table*}[tbp]
\centering\scriptsize
\caption{Evaluation overview. Each row measures one value of one dimension from Table~\ref{tab:taxonomy} on a representative protocol. \emph{Base} and \emph{attack} are percentages and $\Delta$ is their difference in points, except on the realized-MEV rows, where they are MEV amounts and $\Delta$ is a relative change.
}
\label{tab:cells}
\setlength{\tabcolsep}{3pt}
\renewcommand{\arraystretch}{0.88}
\begin{tabular}{@{}llllrrclc@{}}
\toprule
\textbf{ID} & \textbf{Value} & \textbf{Primitive} & \textbf{Protocol} & \textbf{Base} & \textbf{Attack} &
\textbf{$\Delta$} & \textbf{Structural cause} & \textbf{Metric} \\
\midrule
\multicolumn{9}{@{}l}{\textbf{A \; Adversary}}\\
\rowcolor{Headline}
\multicolumn{9}{@{}l}{\quad\emph{A\ref{dim:action} Action}}\\
A1.1 & front-run & fissure & Narwhal-Tusk & $52.85$ & $100.0$ & $+47.15$ & certificate starvation & all-pairs \\
A1.2 & back-run & sluggish & AlephBFT & $0.00$ & $10.06$ & $+10.06$ & reactive path is closed & $L_1$ (immediate back-run) \\
A1.3 & sandwich & fissure & Bullshark & $50.00$ & $81.82$ & $+31.82$ & reaches both sides & triplet sandwich \\
A1.4 & censor & vote-withhold & Bullshark & $0.35$ & $84.40$ & $+84.05$ & vote is the admission gate & inclusion \\
\rowcolor{Headline}
\multicolumn{9}{@{}l}{\quad\emph{A\ref{dim:primitive} Primitive}}\\
A2.1 & fissure & fissure & Narwhal-Tusk & $52.85$ & $100.0$ & $+47.15$ & admission gate & all-pairs \\
A2.2 & speculative & speculative & AlephBFT & $49.94$ & $96.39$ & $+46.45$ & free candidate selection & all-pairs \\
A2.3 & sluggish & sluggish & Mahi-Mahi & $50.01$ & $80.95$ & $+30.94$ & round-first ordering & all-pairs \\
A2.4 & silent-exc-leader & silent-exc-leader & Mysticeti & $56.30$ & $76.09$ & $+19.79$ & leader heads its commit & all-pairs \\
A2.5 & SLW & SLW & Mysticeti & $210.80$ & $359.80$ & $+70.7\%$ & forfeits one anchor & realized MEV \\
A2.6 & proposal-timestamp & proposal-timestamp & Mysticeti & $56.30$ & $67.20$ & $+10.90$ & no upper bound on stamps & all-pairs \\
A2.7 & LVW & LVW & Mysticeti & $87.40$ & $88.70$ & $+1.30$ & four blames, nine needed & same-round \\
A2.8 & vote-withhold & vote-withhold & Bullshark & $0.35$ & $84.40$ & $+84.05$ & no certificate ever forms & inclusion \\
\rowcolor{Headline}
\multicolumn{9}{@{}l}{\quad\emph{A\ref{dim:relation} Relationship}}\\
A3.1 & solo & silent-exc-leader & Mysticeti & $56.30$ & $75.00$ & $+18.70$ & one validator suffices & all-pairs \\
A3.2 & competing & silent-exc-leader & Mysticeti & $56.30$ & $76.09$ & $+19.79$ & independents starve DAG & all-pairs \\
A3.3 & colluding & fissure & Bullshark & $50.00$ & $68.25$ & $+18.25$ & exclusion concentrates & committing-height \\
A3.4 & multi-group & fissure & Bullshark & $50.00$ & $57.89$ & $+7.89$ & fragmenting dilutes the gain & committing-height \\
\rowcolor{Headline}
\multicolumn{9}{@{}l}{\quad\emph{A\ref{dim:info} Information}}\\
A4.1 & local & fissure & Narwhal-Tusk & $52.85$ & $100.0$ & $+47.15$ & own view suffices & all-pairs \\
A4.2 & global & speculative & AlephBFT & $0.29$ & $77.78$ & $+77.49$ & winning rank is knowable & same-round \\
A4.3 & oracle & speculative & Mahi-Mahi & $50.01$ & $95.92$ & $+45.91$ & no mempool, oracle-fed & all-pairs \\
\rowcolor{Headline}
\multicolumn{9}{@{}l}{\quad\emph{A\ref{dim:incentive} Incentive}}\\
A5.1 & none & fissure & Bullshark & $50.00$ & $38.10$ & $-11.90$ & no accomplice, self-cost & committing-height \\
A5.2 & bribery & fissure & Bullshark & $50.00$ & $66.29$ & $+16.29$ & coalition can be bought & committing-height \\
\rowcolor{Headline}
\multicolumn{9}{@{}l}{\quad\emph{A\ref{dim:fraction} Fraction}}\\
A6.1 & $\alpha{=}1/13$ & fissure & Bullshark & $50.00$ & $38.10$ & $-11.90$ & self-cost, victim unharmed & committing-height \\
A6.2 & $\alpha{=}4/13$ & fissure & Bullshark & $50.00$ & $58.21$ & $+8.21$ & refusals outweigh self-cost & committing-height \\
A6.3 & $\alpha{=}6/13$ & fissure & Bullshark & $50.00$ & $68.25$ & $+18.25$ & refusals compound on one victim & committing-height \\
\midrule
\multicolumn{9}{@{}l}{\textbf{P \; Protocol}}\\
\rowcolor{Headline}
\multicolumn{9}{@{}l}{\quad\emph{P\ref{dim:type} DAG type}}\\
P1.1 & certified & fissure & Narwhal-Tusk & $52.85$ & $100.0$ & $+47.15$ & quorum admission gate & all-pairs \\
P1.2 & uncertified & fissure & Mysticeti & $50.00$ & $54.30$ & $+4.30$ & admitted on arrival & all-pairs \\
\rowcolor{Headline}
\multicolumn{9}{@{}l}{\quad\emph{P\ref{dim:ordering} Ordering rule}}\\
P2.1 & leader-anchored & silent-exc-leader & Mysticeti & $56.30$ & $76.09$ & $+19.79$ & leader heads the batch & all-pairs \\
P2.2 & round-only & sluggish & Bullshark & $51.86$ & $95.80$ & $+43.94$ & round sort rewards delay & all-pairs \\
P2.3 & slot-based & fissure & Autobahn & $50.08$ & $51.39$ & $+1.31$ & no shared seam & all-pairs \\
P2.4 & election-based & speculative & AlephBFT & $49.94$ & $96.39$ & $+46.45$ & grindable hash election & all-pairs \\
\rowcolor{Headline}
\multicolumn{9}{@{}l}{\quad\emph{P\ref{dim:tiebreak} Tiebreak}}\\
P3.1 & author & none & Mysticeti & $50.00$ & $56.30$ & $+6.30$ & index buys priority & all-pairs \\
P3.2 & round & none & Bullshark & $50.00$ & $49.98$ & $-0.02$ & no identity in the sort key & all-pairs \\
P3.3 & digest & none & Mysticeti & $56.30$ & $54.29$ & $-2.01$ & removes part of the tax & all-pairs \\
P3.4 & seeded & none & Mysticeti & $56.30$ & $54.10$ & $-2.20$ & unpredictable tie order & all-pairs \\
P3.5 & order-fair & none & Mysticeti & $56.30$ & $52.50$ & $-3.80$ & identity-free ordering & all-pairs \\
\rowcolor{Headline}
\multicolumn{9}{@{}l}{\quad\emph{P\ref{dim:leader} Leader rule}}\\
P4.1 & rotation & silent-exc-leader & Mysticeti & $56.30$ & $76.09$ & $+19.79$ & predictable anchors & all-pairs \\
P4.2 & stake & silent-exc-leader & Mysticeti & $55.70$ & $73.10$ & $+17.40$ & weight buys anchors & all-pairs \\
P4.3 & reputation & silent-exc-leader & Mysticeti & $53.50$ & $52.10$ & $-1.40$ & scoring resists gaming & all-pairs \\
\rowcolor{Headline}
\multicolumn{9}{@{}l}{\quad\emph{P\ref{dim:tuning} Tuning}}\\
P5.1 & protocol-specific & speculative & Mahi-Mahi & $51.98$ & $83.33$ & $+31.35$ & longer wave, more rounds to grind & same-round \\
\midrule
\multicolumn{9}{@{}l}{\textbf{T \; Target}}\\
\rowcolor{Headline}
\multicolumn{9}{@{}l}{\quad\emph{T\ref{dim:victim} Victims}}\\
T1.1 & single & fissure & Bullshark & $50.00$ & $66.29$ & $+16.29$ & undiluted budget & committing-height \\
T1.2 & multiple & fissure & Bullshark & $50.00$ & $38.10$ & $-11.90$ & budget splits & committing-height \\
\rowcolor{Headline}
\multicolumn{9}{@{}l}{\quad\emph{T\ref{dim:metric} Metric}}\\
T2.1 & all-pairs & fissure & Bullshark & $51.86$ & $49.96$ & $-1.90$ & global position & all-pairs \\
T2.2 & same-round & fissure & Bullshark & $53.26$ & $63.38$ & $+10.12$ & round-level race & same-round \\
T2.3 & committing-height & fissure & Bullshark & $50.00$ & $66.29$ & $+16.29$ & targeted pair only & committing-height \\
T2.4 & realized MEV & silent-exc-leader & Mysticeti & $58.00$ & $19.00$ & $-67.2\%$ & value-weighted capture & realized MEV \\
\rowcolor{Headline}
\multicolumn{9}{@{}l}{\quad\emph{T\ref{dim:value} Value}}\\
T3.1 & uniform & silent-exc-leader & Mysticeti & $58.00$ & $19.00$ & $-67.2\%$ & flat value model & realized MEV \\
T3.2 & pareto & silent-exc-leader & Mysticeti & $217.30$ & $57.10$ & $-73.7\%$ & heavy tail concentrates & realized MEV \\
T3.3 & lognormal & silent-exc-leader & Mysticeti & $2269.20$ & $477.30$ & $-79.0\%$ & heaviest tail & realized MEV \\
\midrule
\multicolumn{9}{@{}l}{\textbf{D \; Deployment}}\\
\rowcolor{Headline}
\multicolumn{9}{@{}l}{\quad\emph{D\ref{dim:size} Size $n$}}\\
D1.1 & $n{=}13$ & silent-exc-leader & Mysticeti & $56.30$ & $76.10$ & $+19.80$ & anchor share $\propto 1/n$ & all-pairs \\
D1.2 & $n{=}25$ & silent-exc-leader & Mysticeti & $54.80$ & $59.20$ & $+4.40$ & anchor share $\propto 1/n$ & all-pairs \\
D1.3 & $n{=}49$ & silent-exc-leader & Mysticeti & $49.90$ & $62.90$ & $+13.00$ & control reaches fair line & all-pairs \\
\rowcolor{Headline}
\multicolumn{9}{@{}l}{\quad\emph{D\ref{dim:stake} Stake}}\\
D2.1 & equal & silent-exc-leader & Mysticeti & $55.70$ & $75.40$ & $+19.70$ & uniform voting weight & all-pairs \\
D2.2 & skewed & silent-exc-leader & Mysticeti & $55.70$ & $73.10$ & $+17.40$ & weight buys no position & all-pairs \\
\rowcolor{Headline}
\multicolumn{9}{@{}l}{\quad\emph{D\ref{dim:geo} Geo-distribution}}\\
D3.1 & LAN & silent-exc-leader & Mysticeti & $55.71$ & $71.67$ & $+15.96$ & deterministic order & all-pairs \\
D3.2 & WAN & silent-exc-leader & Mysticeti & $55.71$ & $71.67$ & $+15.96$ & geo-invariant order & all-pairs \\
D3.3 & asymmetric & silent-exc-leader & Mysticeti & $55.71$ & $75.40$ & $+19.69$ & geo-invariant order & all-pairs \\
\bottomrule
\end{tabular}
\end{table*}

Our evaluation measures the impact of MEV attacks on DAG-based BFT protocols through experiments introduced in \S\ref{sec:space}: one per dimension value, each instantiated on whichever candidate protocol's structure makes that value feasible. We instrument six production DAG-based BFT protocols (Narwhal-Tusk, Bullshark, Mysticeti, AlephBFT, Mahi-Mahi, and Autobahn), arm one attacker behavior at a time through an environment-gated hook, and record the commit order.

\ifextend
Every experiment consists of two sets of runs: attack runs and control runs, the same experiment with the hook disarmed, scored under the paired-control discipline of \S\ref{sec:method}. All reported experiments were executed on dedicated CloudLab nodes (hardware type \texttt{c6525-25g}: 16-core AMD 7302P CPU, 128GB ECC Memory, 25Gb Ethernet) \cite{duplyakin2019design}.
\fi

Table~\ref{tab:cells} summarizes the results. Each row measures one value of one dimension from Table~\ref{tab:taxonomy} on a representative protocol. We fix a default value for every dimension in the adversary (excluding primitive), target (excluding metric), and deployment families, and use it in every experiment unless an experiment states otherwise. Protocol-family dimensions have no default, since we run experiments on different representative protocols. The primitive and metric dimensions are excluded, as the protocol and action jointly determine which primitive and metric apply. The default configuration is:
[A\ref{dim:action}: {\sf front-run}, A\ref{dim:relation}: {\sf solo}, A\ref{dim:info}: {\sf local}, A\ref{dim:incentive}: {\sf none}, A\ref{dim:fraction}: $4/13$, T\ref{dim:victim}: {\sf single}, T\ref{dim:value}: {\sf uniform}, D\ref{dim:size}: $13$, D\ref{dim:stake}: {\sf equal}, D\ref{dim:geo}: {\sf LAN}].

The six protocols take the following P dimension values (type, ordering rule, tiebreak, leader rule).
\emph{Narwhal-Tusk}: {\sf certified}, {\sf leader-anchored}, ties by {\sf digest}, leader by random {\sf rotation}.
\emph{Bullshark}: {\sf certified}, {\sf leader-anchored}, ties by {\sf round}, leader by fixed {\sf rotation}.
\emph{Mysticeti}: {\sf uncertified}, {\sf leader-anchored}, ties by {\sf author}, leader by fixed {\sf rotation}.
\emph{AlephBFT}: {\sf certified}, {\sf election-based}; its election sorts a rounds candidate blocks by hash and rotates that order by a configured seed to pick the head, so the election is its leader rule and it has no separate tiebreak.
\emph{Mahi-Mahi}: {\sf uncertified}, {\sf leader-anchored}, ties by {\sf round}, several leaders per round by random {\sf rotation}.
\emph{Autobahn}: {\sf certified} (on a proof of availability of $f{+}1$ votes), {\sf slot-based}, order fixed by the lane interleave, leader by fixed {\sf rotation}.
Tuning (P\ref{dim:tuning}) is protocol-specific; we name the parameters we vary alongside that dimension below.
\ifextend No protocol's paper fixes the tiebreak (P\ref{dim:tiebreak}); each says only that any deterministic rule will do, so every value above is the implementation's own choice.\ \fi

The \emph{primitive} column in Table~\ref{tab:cells} names the legal move the attacker makes to reach the position it wants, which is the default {\sf fissure} unless the row says otherwise and is {\sf none} for the rows that measure the protocol with no attack running. The \emph{base} column reports the experiment with the hook disarmed, \emph{attack} the same experiment with it armed, and $\Delta$ the change between them. The \emph{structural cause} is the protocol property that makes that value effective, and \emph{metric} gives the metric used. Rows are not directly comparable, since the protocol and the configuration both change from one row to the next. Within a dimension the picture is tighter: nine of the seventeen dimensions vary their value inside a single implementation, so the value is the only thing that moves, and three more do so with one cross-protocol check. The remaining five are cross-protocol by necessity. For DAG type (P\ref{dim:type}) and the ordering rule (P\ref{dim:ordering}) the value \emph{is} the protocol, so those rest on comparison across implementations rather than a switch thrown inside one; action (A\ref{dim:action}), primitive (A\ref{dim:primitive}) and information (A\ref{dim:info}) span protocols to establish breadth rather than mechanism.
\ifextend The default configuration is itself a measured cell: fissure on Narwhal-Tusk takes the rate from $52.85$ to $100.0$, a lift of $+47.2$, and that single experiment carries the default value of four dimensions at once (\cellref{A1.1}, \cellref{A2.1}, \cellref{A4.1}, \cellref{P1.1}).\ \fi
A single experiment fixes one value in every dimension at once, so one experiment might answer several rows, e.g., \cellref{A2.4} and \cellref{A3.2}. Two rows depart from the defaults above: the coalition rows \cellref{A3.3} and \cellref{A3.4} run at $\alpha{=}6/13$ rather than $4/13$, since a coalition must be large enough to be worth splitting, so \cellref{A3.3} is the same experiment as \cellref{A6.3}.
\ifextend The silent-except-leader experiment on Mysticeti, for instance, also fixes a leader-anchored ordering rule and a round-robin leader schedule, and so reports the same $56.30 \rightarrow 76.09$ in \cellref{A2.4}, \cellref{A3.2}, \cellref{P2.1} and \cellref{P4.1}. Two further pairs work the same way: \cellref{A2.2} and \cellref{P2.4} are one speculative experiment on AlephBFT, and \cellref{A3.3} and \cellref{A6.3} are one six-attacker run on Bullshark.
We read such runs along each axis instead of repeating them, because re-running the same configuration under a second heading would return the same numbers at extra cost and would count one measurement as several results. The table therefore contains fewer independent measurements than it has rows.\ \fi

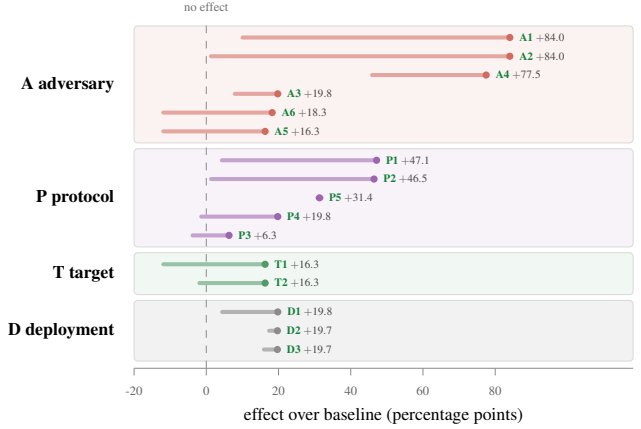
\begin{figure}[tbp]
\centering
\resizebox{\columnwidth}{!}{
\begin{tikzpicture}[font=\scriptsize, x=0.050cm, y=1cm,
 band/.style={rounded corners=1.5pt, draw=black!15},
 seg/.style={line width=1.6pt, line cap=round},
 dot/.style={circle, inner sep=1.0pt},
 fam/.style={font=\scriptsize\bfseries, anchor=east},
 lab/.style={font=\fontsize{4.4}{5}\selectfont, text=black!72, inner sep=1pt}]
\node[band, fill=AttackFissure!6, minimum width=6.91cm, minimum height=1.62cm] at (49.1,4.650) {};
\node[fam] at (-23.0,4.650){A adversary};
\node[band, fill=AttackSpeculative!6, minimum width=6.91cm, minimum height=1.36cm] at (49.1,3.080) {};
\node[fam] at (-23.0,3.080){P protocol};
\node[band, fill=AttackSilent!6, minimum width=6.91cm, minimum height=0.58cm] at (49.1,2.030) {};
\node[fam] at (-23.0,2.030){T target};
\node[band, fill=black!5, minimum width=6.91cm, minimum height=0.84cm] at (49.1,1.240) {};
\node[fam] at (-23.0,1.240){D deployment};
\draw[black!45, dashed] (0,0.72) -- (0,5.54);
\node[font=\tiny, text=black!55, anchor=south] at (0,5.54) {no effect};
\draw[seg,AttackFissure!45](10.06,5.300)--(84.05,5.300); \node[dot,fill=AttackFissure!75]at(84.05,5.300){}; \node[lab,right]at(85.85,5.300){\dimref{A1} $+84.0$};
\draw[seg,AttackFissure!45](1.30,5.040)--(84.05,5.040); \node[dot,fill=AttackFissure!75]at(84.05,5.040){}; \node[lab,right]at(85.85,5.040){\dimref{A2} $+84.0$};
\draw[seg,AttackFissure!45](45.91,4.780)--(77.49,4.780); \node[dot,fill=AttackFissure!75]at(77.49,4.780){}; \node[lab,right]at(79.29,4.780){\dimref{A4} $+77.5$};
\draw[seg,AttackFissure!45](7.89,4.520)--(19.79,4.520); \node[dot,fill=AttackFissure!75]at(19.79,4.520){}; \node[lab,right]at(21.59,4.520){\dimref{A3} $+19.8$};
\draw[seg,AttackFissure!45](-11.90,4.260)--(18.25,4.260); \node[dot,fill=AttackFissure!75]at(18.25,4.260){}; \node[lab,right]at(20.05,4.260){\dimref{A6} $+18.3$};
\draw[seg,AttackFissure!45](-11.90,4.000)--(16.29,4.000); \node[dot,fill=AttackFissure!75]at(16.29,4.000){}; \node[lab,right]at(18.09,4.000){\dimref{A5} $+16.3$};
\draw[seg,AttackSpeculative!45](4.30,3.600)--(47.15,3.600); \node[dot,fill=AttackSpeculative!75]at(47.15,3.600){}; \node[lab,right]at(48.95,3.600){\dimref{P1} $+47.1$};
\draw[seg,AttackSpeculative!45](1.31,3.340)--(46.45,3.340); \node[dot,fill=AttackSpeculative!75]at(46.45,3.340){}; \node[lab,right]at(48.25,3.340){\dimref{P2} $+46.5$};
\draw[seg,AttackSpeculative!45](30.75,3.080)--(31.95,3.080); \node[dot,fill=AttackSpeculative!75]at(31.35,3.080){}; \node[lab,right]at(33.15,3.080){\dimref{P5} $+31.4$};
\draw[seg,AttackSpeculative!45](-1.40,2.820)--(19.79,2.820); \node[dot,fill=AttackSpeculative!75]at(19.79,2.820){}; \node[lab,right]at(21.59,2.820){\dimref{P4} $+19.8$};
\draw[seg,AttackSpeculative!45](-3.80,2.560)--(6.30,2.560); \node[dot,fill=AttackSpeculative!75]at(6.30,2.560){}; \node[lab,right]at(8.10,2.560){\dimref{P3} $+6.3$};
\draw[seg,AttackSilent!45](-11.90,2.160)--(16.29,2.160); \node[dot,fill=AttackSilent!75]at(16.29,2.160){}; \node[lab,right]at(18.09,2.160){\dimref{T1} $+16.3$};
\draw[seg,AttackSilent!45](-1.90,1.900)--(16.29,1.900); \node[dot,fill=AttackSilent!75]at(16.29,1.900){}; \node[lab,right]at(18.09,1.900){\dimref{T2} $+16.3$};
\draw[seg,black!30](4.40,1.500)--(19.80,1.500); \node[dot,fill=black!45]at(19.80,1.500){}; \node[lab,right]at(21.60,1.500){\dimref{D1} $+19.8$};
\draw[seg,black!30](17.40,1.240)--(19.70,1.240); \node[dot,fill=black!45]at(19.70,1.240){}; \node[lab,right]at(21.50,1.240){\dimref{D2} $+19.7$};
\draw[seg,black!30](15.96,0.980)--(19.69,0.980); \node[dot,fill=black!45]at(19.69,0.980){}; \node[lab,right]at(21.49,0.980){\dimref{D3} $+19.7$};
\draw[black!45] (-20.0,0.72)--(118.2,0.72);
\foreach \x in {-20,0,20,40,60,80}
  {\draw[black!45](\x,0.72)--(\x,0.62); \node[font=\tiny,text=black!60,anchor=north]at(\x,0.60){\x};}
\node[font=\scriptsize, anchor=north] at (49.1,0.30){effect over baseline (percentage points)};
\end{tikzpicture}}
\caption{One line per \emph{dimension}, arranged by family and sorted by effect within each family. Each value is scored on its own metric against its own baseline, so a line shows the spread a dimension's values produced, not a ranking between dimensions. The realized-MEV rows are relative changes rather than point differences and cannot be placed on this axis, so \dimref{T3}, \cellref{A2.5} and \cellref{T2.4} are omitted.
\ifextend The dot marks the value furthest from no effect, and a span crosses metrics wherever a dimension's values are scored differently. \dimref{A1} is the clearest case: its line runs from the $L_1$ back-run rate at $+10.1$ (\cellref{A1.2}) to the inclusion rate at $+84.1$ (\cellref{A1.4}), passing through all-pairs and the triplet sandwich, so its width is partly the four actions and partly the four metrics applied to them.\fi}
\label{fig:designspace}
\end{figure}

Figure~\ref{fig:designspace} condenses the same table into one line per dimension. Each line spans the range of effects produced by that dimension's values: a long line indicates a dimension that strongly influences outcomes. Because each value is scored on its own metric, a single line can span multiple metrics, so the figure serves as a guide to which dimensions merit closer attention rather than a precise quantitative ranking.

\ifextend The remainder of this section discusses the dimensions in the order given by Table~\ref{tab:taxonomy}.\ \fi

\subsection{The Adversary Family}\label{secx:adversary}

\begin{adver}
\label{sec:action}
\textbf{Action.}
We extend the \dimref{A1} row of Table~\ref{tab:cells} with the breadth campaign of~\cite{mirzaei2027benchmark}, which measures the first three actions' attack success rate (ASR) on four protocols that between them cover four distinct mechanisms for deciding what may follow a committed block: certified branch ordering (Bullshark), uncertified wave-based ordering (Mysticeti), leaderless virtual voting (AlephBFT), and slot-based assembly (Autobahn). Figure~\ref{fig:action} reports the results. \ifextend An action only succeeds if some primitive can reach the position it aims at, so where the default primitive of Table~\ref{tab:cells}, fissure, cannot reach it, we substitute one that can and say which, and why, at that row. Front-running (\cellref{A1.1}) and sandwiching (\cellref{A1.3}) use fissure; the other two do not.\ \fi

Two findings stand out.
First, a protocol's vulnerability varies across attack types. For example, Bullshark is more vulnerable to front-running than to back-running $L_2$ ($94\%$ vs. $67\%$), whereas Mysticeti is more vulnerable to back-running $L_2$ than to front-running ($63\%$ vs. $54\%$).
Second, how an attack is defined matters. For back-running, placing a block anywhere after the victim (cumulative) is easy. Placing it \emph{immediately} after the victim ($L_1$), however, is hard, since that single position is contested by every honest block that also references the victim. This gap is evident when comparing the cumulative success rate to the $L_1$ one for back-running: Bullshark falls from $98\%$ to $33\%$, Mysticeti from $100\%$ to $35\%$, and Autobahn from $52\%$ to $8\%$.
Sandwiching is harder still, since it requires succeeding on both sides at once\ifextend, so its success rate tracks whichever side the protocol makes harder for the attacker\fi. Bullshark, however, is still vulnerable to sandwiching, taking the triplet sandwich rate from $50.00$ to $81.82$, a lift of $+31.8$ (\cellref{A1.3}), because certification, the admission gate defined in P\ref{dim:type}, gives the attacker the same leverage on the front side that it already has on the back.

\ifextend Under fissure, AlephBFT's back-run rate is $0$ at both $L_1$ and $L_2$, and only $7.7\%$ cumulatively.  Fissure denies a victim the support it needs, but it doesn't hand the attacker the position right behind that victim, and on AlephBFT nothing else fills that gap either. What does fill it is a primitive that lets the attacker control its own timing: under sluggish, the same protocol jumps to $10.06\%$ at $L_1$ (\cellref{A1.2}), which is why that row is the one exception to the default primitive. So back-running turns out to depend less on the attack itself than on whether the primitive can actually put the attacker where the attack needs it to be.\ \fi

The last action is censorship, which we measure by \emph{inclusion}: how far the victim falls behind the other honest validators in committed-block count. \ifextend By this measure, fissure censors nothing under any protocol. It does refuse to reference the victim, but a parent reference and an admission vote are different acts: fissure leaves the victim out of the \emph{attacker's own} parent set, while the certificate that lets the victim's block count is assembled from signatures on the victim's \emph{header}, which the attacker is never asked to supply. Refusing to point at a block is therefore not refusing to admit it, and the block still commits. Increasing the number of validators that omit the victim from their parent sets, from four up to seven, still leaves the victim committing about as often as everyone else, with inclusion gaps of only $-0.5\%$ to $0.0\%$.\ \fi
Censorship comes from withholding \emph{certification}, \ifextend and this is the second departure from the default primitive: fissure withholds a parent reference, which moves a block, whereas censorship keeps a block out of the order altogether,\ \fi e.g., on Bullshark, when four ($f$) validators withhold their votes on the victim's headers, $84.4\%$ of the victim's blocks never commit, compared to a $0.35\%$ baseline (\cellref{A1.4}, \cellref{A2.8}). This is because committing a block requires $2f{+}1 = 9$ votes. With four validators refusing, only nine willing voters remain, so the victim needs every one of them to vote in every round, and ordinary asynchrony denies it that unanimity in most rounds.
\ifextend The implication for protocol design is that repositioning and exclusion are distinct capabilities. Certification is what makes both available in the same protocol: a design that admits blocks on arrival, without a certification gate, stops both attacks at once.\fi
\end{adver}

\definecolor{AttackFissure}{RGB}{214,39,40}
\definecolor{AttackSpeculative}{RGB}{148,103,189}
\definecolor{AttackSluggish}{RGB}{31,119,180}

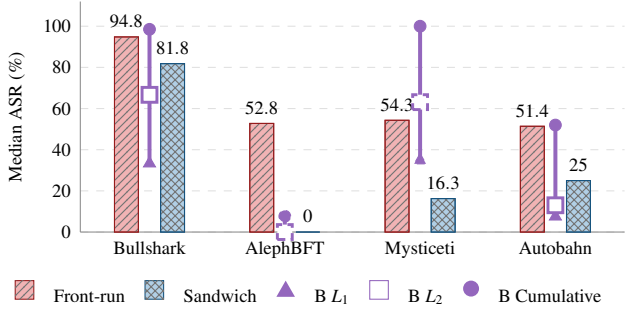
\begin{figure}[t]
\begin{tikzpicture}
\begin{axis}[
  scale only axis,
  width=0.85\linewidth,
  height=0.36\linewidth,
  ymin=0, ymax=112,
  xmin=0.55, xmax=4.09,
  ytick={0,20,40,60,80,100},
  yticklabel style={font=\scriptsize},
  ylabel={Median ASR (\%)},
  ylabel style={font=\scriptsize},
  xtick={1.00,1.88,2.76,3.64},
  xticklabels={Bullshark,AlephBFT,Mysticeti,Autobahn},
  x tick label style={font=\scriptsize, align=center},
  ymajorgrids, grid style={dashed,black!12},
  tick style={draw=none},
  axis line style={draw=black!45},
  axis x line*=bottom,
  axis y line*=left,
  clip=false,
]

\addplot+[
  ybar, bar width=9pt, mark=none,
  fill=AttackFissure!35, draw=AttackFissure!70!black,
  postaction={pattern=north east lines, pattern color=black!55},
  nodes near coords,
  every node near coord/.append style={font=\scriptsize, anchor=south, text=black, /pgf/number format/.cd, fixed, precision=1},
] coordinates {(0.85,94.80) (1.73,52.77) (2.61,54.30) (3.49,51.39)};

\addplot+[
  ybar, bar width=9pt, mark=none,
  fill=AttackSluggish!35, draw=AttackSluggish!70!black,
  postaction={pattern=crosshatch, pattern color=black!55},
  nodes near coords,
  every node near coord/.append style={font=\scriptsize, anchor=south, text=black, /pgf/number format/.cd, fixed, precision=1},
] coordinates {(1.15,81.82) (2.03,0.00) (2.91,16.26) (3.79,25.00)};

\draw[AttackSpeculative, line width=1.5pt] (axis cs:1.00,33.33) -- (axis cs:1.00,98.50);
\addplot+[only marks, mark=triangle*, mark options={fill=AttackSpeculative}, color=AttackSpeculative, mark size=2.6pt] coordinates {(1.00,33.33)};
\addplot+[only marks, mark=square*, mark options={fill=white}, color=AttackSpeculative, mark size=2.8pt, line width=1.1pt] coordinates {(1.00,66.67)};
\addplot+[only marks, mark=*, mark options={fill=AttackSpeculative}, color=AttackSpeculative, mark size=2.2pt] coordinates {(1.00,98.50)};
\draw[AttackSpeculative, line width=1.5pt] (axis cs:1.88,0.00) -- (axis cs:1.88,7.68);
\addplot+[only marks, mark=triangle*, mark options={fill=AttackSpeculative}, color=AttackSpeculative, mark size=2.6pt] coordinates {(1.88,0.00)};
\addplot+[only marks, mark=square*, mark options={fill=white}, color=AttackSpeculative, mark size=2.8pt, line width=1.1pt] coordinates {(1.88,0.00)};
\addplot+[only marks, mark=*, mark options={fill=AttackSpeculative}, color=AttackSpeculative, mark size=2.2pt] coordinates {(1.88,7.68)};
\draw[AttackSpeculative, line width=1.5pt] (axis cs:2.76,34.81) -- (axis cs:2.76,100.00);
\addplot+[only marks, mark=triangle*, mark options={fill=AttackSpeculative}, color=AttackSpeculative, mark size=2.6pt] coordinates {(2.76,34.81)};
\addplot+[only marks, mark=square*, mark options={fill=white}, color=AttackSpeculative, mark size=2.8pt, line width=1.1pt] coordinates {(2.76,63.11)};
\addplot+[only marks, mark=*, mark options={fill=AttackSpeculative}, color=AttackSpeculative, mark size=2.2pt] coordinates {(2.76,100.00)};
\draw[AttackSpeculative, line width=1.5pt] (axis cs:3.64,7.53) -- (axis cs:3.64,51.98);
\addplot+[only marks, mark=triangle*, mark options={fill=AttackSpeculative}, color=AttackSpeculative, mark size=2.6pt] coordinates {(3.64,7.53)};
\addplot+[only marks, mark=square*, mark options={fill=white}, color=AttackSpeculative, mark size=2.8pt, line width=1.1pt] coordinates {(3.64,13.01)};
\addplot+[only marks, mark=*, mark options={fill=AttackSpeculative}, color=AttackSpeculative, mark size=2.2pt] coordinates {(3.64,51.98)};

\end{axis}

\matrix[anchor=north, column sep=0.14cm, row sep=0pt, font=\scriptsize] at ($(current bounding box.south)+(0,-0.05cm)$) {
  \node[draw=AttackFissure!70!black, fill=AttackFissure!35, minimum width=7pt, minimum height=7pt, inner sep=0pt, postaction={pattern=north east lines, pattern color=black!55}] {}; & \node {Front-run}; &
  \node[draw=AttackSluggish!70!black, fill=AttackSluggish!35, minimum width=7pt, minimum height=7pt, inner sep=0pt, postaction={pattern=crosshatch, pattern color=black!55}] {}; & \node {Sandwich}; &
  \node[inner sep=0pt] {\tikz[baseline=-0.6ex]\fill[AttackSpeculative] (0pt,0pt) -- (7pt,0pt) -- (3.5pt,6.1pt) -- cycle;}; & \node {B $L_1$}; &
  \node[draw=AttackSpeculative, minimum width=7pt, minimum height=7pt, inner sep=0pt, fill=white] {}; & \node {B $L_2$}; &
  \node[inner sep=0pt] {\tikz[baseline=-0.6ex]\fill[AttackSpeculative] (3.5pt,3.5pt) circle (3.0pt);}; & \node {B Cumulative}; \\
};
\end{tikzpicture}
\caption{Fissure attack success rate under different Actions}
  \label{fig:action}
\end{figure}

\begin{adver}
\label{sec:primitive}
\textbf{Primitive.}
The first three primitives are more general, and are measured on all six protocols in that same campaign~\cite{mirzaei2027benchmark}, as reported in Figure~\ref{fig:primitive}.
Fissure leaves the victim out of the parent set it references, so it needs an admission gate to hold shut: on Narwhal-Tusk it improves the success rate to $100.0$ (\cellref{A2.1}).
Speculative builds several valid candidate blocks and publishes whichever sorts best. On AlephBFT, it takes the success rate to $96.39$ (\cellref{A2.2}) because AlephBFT lets validators choose their own block contents.
Sluggish holds its own proposal back into a more valuable round. On Mahi-Mahi, it improves the rate to $80.95$ (\cellref{A2.3}) because Mahi-Mahi sorts by round first.

Each of the three primitives needs one property of the protocol and gets nothing without it.
If a block must gather a certificate before it counts, e.g., Narwhal-Tusk and Bullshark, take \emph{fissure}.
If blocks are admitted on arrival, the ordering rule becomes important. When a separate election decides which block leads, e.g., AlephBFT and Mahi-Mahi, take \emph{speculative}, and when the order sorts by round before anything else, e.g., Mahi-Mahi, take \emph{sluggish}.
We discuss Mysticeti's vulnerabilities in detail below. Autobahn, meanwhile, sits close to the fair line under all three primitives, because a slot-based rule assembles each slot from every validator's own chain, so there is no shared seam for a proposer to fight over and no property for any of these primitives to consume.

Four of the remaining five are measured on Mysticeti, and each one turns a different piece of its machinery against it; the fifth, vote-withholding, needs a certificate to deny and so is measured on Bullshark.
Silent-except-leader exploits the leader-anchored commit rule: on Mysticeti, a block a validator proposes during its own leader round anchors that round's commit, leading to a $19.8$-point ASR rise (\cellref{A2.4}).
Strategic leader withholding (SLW) skips the attacker's own leader round, which folds that wave into the next leader's history and moves the boundary between two commits. It keeps every block it would otherwise have published, so a positional metric records almost nothing, and we measure it using realized MEV instead, under the heavy-tailed value model rather than the uniform default, where it improves the ASR on Mysticeti by $70.7\%$ (\cellref{A2.5}).
Proposal-timestamp stamps a block with a favorable time, and it improves ASR by $10.9$ points because nothing in Mysticeti bounds that timestamp (\cellref{A2.6}).
Leader-vote withholding (LVW) withholds the parent reference a leader needs in order to commit. This attack does not necessarily work:  even if all $f$ attackers withhold their votes, the leader can still receive the $2f+1$ required votes from the remaining validators. It can, however, make the leader's commit more fragile (\cellref{A2.7}).
\end{adver}

\definecolor{AttackFissure}{RGB}{214,39,40}
\definecolor{AttackSpeculative}{RGB}{148,103,189}
\definecolor{AttackSluggish}{RGB}{31,119,180}
\definecolor{AttackBase}{RGB}{127,127,127}

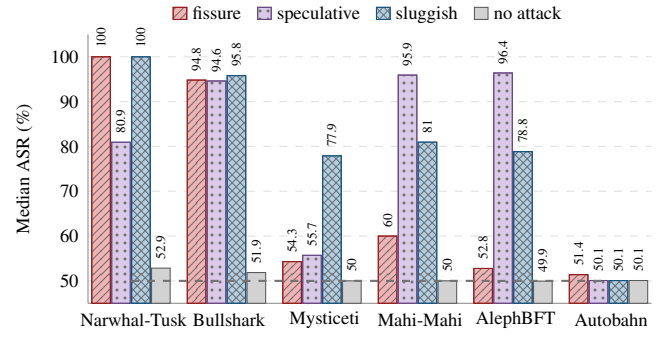
\begin{figure}[t]
\begin{tikzpicture}
\begin{axis}[
  scale only axis,
  width=0.88\linewidth,
  height=0.42\linewidth,
  ybar=0.5pt,
  bar width=7pt,
  enlarge x limits=0.09,
  ymin=45, ymax=105,
  ytick={50,60,70,80,90,100},
  clip=false,
  yticklabel style={font=\scriptsize},
  ylabel={Median ASR (\%)},
  ylabel style={font=\scriptsize},
  symbolic x coords={Narwhal,Bullshark,Mysticeti,Mahi-Mahi,AlephBFT,Autobahn},
  xtick=data,
  xticklabels={Narwhal-Tusk,Bullshark,Mysticeti,Mahi-Mahi,AlephBFT,Autobahn},
  xticklabel style={font=\scriptsize, rotate=0, anchor=center, yshift=-2pt, xshift=1pt},
  ymajorgrids, grid style={dashed,black!12},
  legend style={
    at={(0.5,1.0)}, anchor=south, legend columns=4, draw=none, fill=none,
    font=\scriptsize, /tikz/every even column/.append style={column sep=4pt},
  },
  legend cell align=left,
  legend image code/.code={\draw[#1] (0cm,-0.06cm) rectangle (0.18cm,0.10cm);},
  tick style={draw=none},
  axis line style={draw=black!45},
  axis x line*=bottom,
  axis y line*=left,
]
\addplot+[
  fill=AttackFissure!35, draw=AttackFissure!70!black,
  postaction={pattern=north east lines, pattern color=black!55},
  nodes near coords,
  every node near coord/.append style={font=\tiny, rotate=90, anchor=west, text=black, /pgf/number format/.cd, fixed, precision=1},
] coordinates
  {(Narwhal,100.00) (Bullshark,94.80)
   (Mysticeti,54.30) (Mahi-Mahi,60.00) (AlephBFT,52.77) (Autobahn,51.39)};
\addplot+[
  fill=AttackSpeculative!35, draw=AttackSpeculative!70!black,
  postaction={pattern=dots, pattern color=black!55},
  nodes near coords,
  every node near coord/.append style={font=\tiny, rotate=90, anchor=west, text=black, /pgf/number format/.cd, fixed, precision=1},
] coordinates
  {(Narwhal,80.93) (Bullshark,94.60)
   (Mysticeti,55.70) (Mahi-Mahi,95.92) (AlephBFT,96.39) (Autobahn,50.08)};
\addplot+[
  fill=AttackSluggish!35, draw=AttackSluggish!70!black,
  postaction={pattern=crosshatch, pattern color=black!55},
  nodes near coords,
  every node near coord/.append style={font=\tiny, rotate=90, anchor=west, text=black, /pgf/number format/.cd, fixed, precision=1},
] coordinates
  {(Narwhal,100.00) (Bullshark,95.80)
   (Mysticeti,77.90) (Mahi-Mahi,80.95) (AlephBFT,78.83) (Autobahn,50.07)};
\addplot+[
  fill=AttackBase!35, draw=AttackBase!70!black,
  nodes near coords,
  every node near coord/.append style={font=\tiny, rotate=90, anchor=west, text=black, /pgf/number format/.cd, fixed, precision=1},
] coordinates
  {(Narwhal,52.85) (Bullshark,51.86)
   (Mysticeti,50.00) (Mahi-Mahi,50.01) (AlephBFT,49.94) (Autobahn,50.08)};
\legend{fissure, speculative, sluggish, no attack}
\draw[dashed, thick, black!55] (axis cs:Narwhal,50) -- (axis cs:Autobahn,50);
\end{axis}
\end{tikzpicture}
\caption{Front-run attack success rate under different primitives}
  \label{fig:primitive}
\end{figure}

\begin{adver}
\label{sec:find:collude}\label{sec:find:coord}
\textbf{Relationship.}
\ifextend Relationship takes four values: {\sf solo}, a single attacker; {\sf competing}, several attackers each working on a different victim; {\sf colluding}, several attackers working on the same victim; and {\sf multi-group}, one coalition split across two victims. Count and coordination are usually varied together, which hides which of the two does the work, so where the definitions allow it, we hold the count fixed and change only the relationship.\ \fi
The {\sf solo} and {\sf competing} attacks run silent-except-leader on Mysticeti, and the {\sf colluding} and {\sf multi-group} attacks run fissure on Bullshark, so the dimension is read on an uncertified and a certified leader-anchored protocol. The two attacks on Bullshark are scored on targeted committing-height, because the question is whether the attackers close on one named victim, and an average taken over every attacker/victim pair (all-pairs) cannot separate the pairs aimed at that victim from the rest.

On Mysticeti, even a single attacker ({\sf solo}) reaches $75\%$ ASR (\cellref{A3.1}) and adding three more attackers that act independently ({\sf competing}), reaches only $76.09$ (\cellref{A3.2}) because each attacker repositions its own blocks and does not gain from the others.
On Bullshark, {\sf colluding} attackers aimed at one victim lift ASR by $18.25$ points (\cellref{A3.3}), while the same coalition split across two victims, {\sf multi-group}, lifts it by only $7.9$ points (\cellref{A3.4}). This is because fissure works by denying a victim support, and {\sf multi-group} splits the attacking power between two victims.
\end{adver}

\begin{adver}
\label{sec:find:info}
\textbf{Information.}
Fissure needs nothing beyond the attacker's {\sf local} view; e.g., on Narwhal-Tusk it reaches $100\%$ ASR with local view (\cellref{A4.1}), so giving it more information changes nothing.
The other two rows therefore run \emph{speculative}.
AlephBFT's leader election takes the candidate units of a round, sorts them by hash, rotates that order left by a seed, and elects whichever candidate the rotation brings to the front. The seed is a configuration constant, chosen once by the operator and not redrawn per round; it fixes how far to rotate, so the winner is the entry at rank $\mathrm{seed} \bmod \ell$ of the hash-sorted candidates, where $\ell$ is the number of candidates in the round. For instance, at seed $123$, which makes rank $6$ the winner ($123 \mod 13 =6$), default speculative wins only $0.29\%$ of same-round elections (\cellref{A4.2}), while an attacker that aims at rank $6$ in the hash space wins $77.78\%$ of them. We score this cell on same-round because the contest is which candidate wins that round's election, so success is by definition a placement inside one round.
Note that $\ell$ is the round's actual candidate count, not the committee size, and rounds frequently carry a lower number of candidates. Knowing the seed is therefore not sufficient: the attacker also needs the round's candidate count, which requires observing the round. Knowing both lets the attacker aim at the correct rank every round. The same attack scored all-pairs rather than same-round reaches $96.39\%$ on AlephBFT (\cellref{A2.2}), since a candidate that loses its own election still lands ahead of most blocks in later rounds.
The attacker might become aware of what a transaction is worth by a channel outside the protocol ({\sf oracle}). For instance, Mahi-Mahi has no public mempool, so speculation there cannot be fed from inside; supplied from outside it moves the rate to $95.92\%$ (\cellref{A4.3}).
\ifextend Information is a dimension in the same sense as the metric and the action. The same deployed code, the same primitive, and the same attacker budget produce $0.3\%$ or $77.8\%$ depending only on what the attacker is assumed to know, and a mitigation evaluated against the uninformed end of that range can look sound while being open at the other. \fi
\end{adver}

\begin{adver}
\label{sec:find:bribe}
\textbf{Incentive.}
The default incentive is {\sf none} where the attacker pays nobody (\cellref{A5.1}).
Here we give the attacker a budget instead of more Byzantine nodes. One Byzantine validator runs the attack, and three \emph{honest} validators are paid to take a single specific legal action on its behalf. We compare against the same lone attacker unaided and against four genuine Byzantine attackers. Both rows run fissure on Bullshark and are scored on targeted committing-height, because a bribe is spent on one named victim, and holding protocol and metric fixed is what lets the two be read against each other directly.
\ifextend  This dimension departs from the default attacker fraction of $4/13$: both rows run a \emph{single} Byzantine validator. The claim being tested is that a budget substitutes for Byzantine nodes, so both rows are scored against the same no-attack control as the rest of the Bullshark rows in Table~\ref{tab:cells}, and the only thing that changes between them is whether the budget is spent. \ \fi

On Bullshark, a single attacker (\cellref{A5.1}) scores $38.10\%$, which is even below the $50\%$ fair ASR. This is because refusing to reference the victim block, while slightly reducing the victim's chance to be included, also makes the attacker's own block less well-connected.
Adding three bribed validators lifts it to $66.3\%$ (\cellref{A5.2}), which also exceeds the $58.21\%$ achieved by four real Byzantine attackers (\cellref{A6.2}).
Fissure needs validators to omit the victim block, and a bribed honest validator does the same, so three purchases substitute for three compromises.
Refusing to reference the victim also leaves the refuser's own block less well connected (as discussed for \cellref{A5.1}). With four Byzantine attackers, all four carry that loss, and the damage affects the attacker's success rate, whereas with one attacker and three bribed validators, the three bribed nodes take the same damage, but their weakened blocks do not count towards the success rate.
\ifextend Interestingly, on Mysticeti, the identical bribe changes the result by $\mathbf{0.0}$ points: the attack stays at the $76.09\%$ that silent-except-leader reaches unaided (\cellref{A2.4}), so the payment buys nothing.\fi
\end{adver}

\begin{adver}
\label{sec:find:fraction}
\textbf{Fraction.}
We fix the relationship to {\sf colluding} (instead of {\sf solo}), because we aim to observe what adding attackers to a coalition buys.
All three rows are scored against the same no-attack control, the protocol's fair $50\%$, so the only thing that moves across them is $\alpha$.
We keep the attackers coordinated on one victim and vary only their share of the committee, $\alpha \in \{1/13, 4/13, 6/13\}$, running fissure on Bullshark throughout. The relationship is held on a single victim, so all three rows keep the targeted committing-height metric of \dimref{A3}
rather than the default all-pairs.
At $\alpha{=}1/13$ it costs its own operator $11.9$ points (\cellref{A6.1}); at $\alpha{=}4/13$ it gains $8.2$ points (\cellref{A6.2}); and at $\alpha{=}6/13$ it gains $18.25$ (\cellref{A6.3}).
The negative effect at $\alpha{=}1/13$ is the self-inflicted cost of solo fissure described in A\ref{sec:find:bribe}. Interestingly, adding more attackers does not help once the fraction is sufficient to stall the protocol. We evaluate this on Mysticeti and on silent-except-leader: moving from four attackers to five takes the rate from $76.09$ to $50.01$, because a silent set that large pushes the number of active proposers below the quorum a round needs to advance.
\end{adver}

\subsection{The Protocol Family}\label{secx:protocol}

\begin{proto}
\label{sec:find:cert}
\textbf{DAG type.}
To evaluate the impact of DAG type, we hold the primitive fixed at the default, fissure, and change only whether a block needs a certificate before it counts, running it on Narwhal-Tusk and Bullshark ({\sf certified}) and on Mysticeti ({\sf uncertified}).
Fissure gains $47.2$ points on Narwhal-Tusk (\cellref{P1.1}) and $42.9$ points on Bullshark ($51.9 \rightarrow 94.8$), the two certified protocols, against $4.3$ points on the {\sf uncertified} Mysticeti (\cellref{P1.2}). These rows come from the breadth campaign of Figure~\ref{fig:primitive}, whose Mysticeti control sits at the fair $50.00$ rather than the $56.30$ of the silence campaign, its attacker not being drawn from the low indices that inherit the P\ref{dim:tiebreak} advantage.
The difference is what it takes for a block to count. On a \emph{certified} DAG a block needs a certificate before later blocks can reference it, and one that few validators referenced is less likely to sit inside the anchor's causal history, so it waits for a later anchor and lands further back. Fissure attackers do this by not referencing the victim block. On an \emph{uncertified} DAG, a block joins the graph once it is broadcast, and any one honest validator that references it carries it into the anchor's causal history. As shown in Figure~\ref{fig:primitive}, Fissure succeeds on the two protocols whose certificate is what admits a block to the DAG, and stays within a few points of the fair line on three of the remaining four; Mahi-Mahi is a partial exception, where it reaches $60.0$ against a $50.0$ control.
\end{proto}

\begin{proto}
\label{sec:find:door}
\textbf{Ordering rule.}
If the uncertified design closes the door on fissure, we ask what it leaves open. We run silent-except-leader, in which the attacker proposes only in rounds where the protocol designates it leader and stays quiet otherwise.
Silent-except-leader gains $19.8$ points on Mysticeti (\cellref{P2.1}) with a single attacker. Mysticeti follows {\sf leader-anchored} rule; a block proposed during your leader round sits at the front of your commit, while a block proposed in any other round is one of many inside somebody else's history. Staying silent means never spending a block on a bad position. The attacker trades volume for placement, and every block it publishes lands at an anchor. Interestingly, the same silent-except-leader attack is not effective ($-0.7\%$) on the Bullshark (certified) protocol, as a committed batch is sorted by
round alone in Bullshark. \ifextend So leading a round does not give the leader's block a privileged position \emph{within} that round. Staying silent therefore sacrifices output, and our silent attacker committed $29\%$ fewer blocks than its control, while buying no placement in return. For \emph{these two attacks}, each protocol is exposed through one door and closed on the other, and the two doors are opened by two different dimensions. Certification opens the first, because an admission gate is something an attacker can hold shut against a victim. A leader-anchored ordering rule opens the second, because it makes the rounds a validator leads worth more than the rounds it does not.\ \fi
The two primitives differ in what they manipulate. Fissure attacks \emph{other validators'} positions, so it needs enough attackers to deny a victim meaningful support. Silent-except-leader repositions \emph{the attacker's own} blocks, so it does not need cooperation.
\ifextend Four attackers acting independently reach $76.09$ (\cellref{A3.2}), barely above the $75.00$ a single attacker reaches (\cellref{A3.1}), because they do not have to agree on anything.\ \fi
A {\sf round-only} rule rewards delay; running sluggish on Bullshark improves the success rate by $43.9$ points (\cellref{P2.2}).
A {\sf slot-based} rule leaves almost nothing to reach, e.g., Autobahn assembles each slot from every validator's own chain, so no proposer shares a seam with another, and fissure moves it only by $1.3$ points (\cellref{P2.3}).
An {\sf election-based} rule moves the contest into the election, where speculative grinds the hash and gains $46.5$ points on AlephBFT (\cellref{P2.4}).
\end{proto}

\begin{proto}
\label{sec:find:tiebreak}
\textbf{Tiebreak.}
The five rows of Tiebreak are read differently from the rest of the table, because this dimension measures a property of the protocol rather than an attack.
\cellref{P3.1} and \cellref{P3.2} report what a shipped tiebreak hands out on its own, so their first column is the fair $50\%$ line rather than a measured control: the {\sf author} rule gives $6.3\%$ away for free on Mysticeti, while the {\sf round} rule leaves Bullshark at $49.98\%$, within a fiftieth of a point of fair. The other three replace the author tiebreak and report what survives, so their second column is the score under the new rule and a \emph{reduction} is the desired outcome: {\sf digest} reduces it by $2.0$ points (\cellref{P3.3}), {\sf seeded} by $2.2$ (\cellref{P3.4}), and {\sf order-fair} by $3.8$ (\cellref{P3.5}). These three are the one place where we hold the implementation fixed and change only the ordering rule, each arm scored against a baseline produced under that same rule, so the rule itself is the independent variable rather than the protocol. Removing identity from the key steadily lowers what the protocol gives away with no attacker running, $56.30 \rightarrow 53.73 \rightarrow 52.66$, and leaves both attacks where they were: silence scores $75.94$, $75.94$ and $75.62$ across the three rules, within a third of a point of itself. A tiebreak reaches the free advantage, not the adversary. \appref{sec:mitigation} gives the full comparison.
The metric here is all-pairs, and the author tiebreak orders only blocks of the same round; cross-round pairs are decided by round number and split evenly. When measuring the success rate using the same-round metric, the validator with index $0$ wins $91.1\%$ of its same-round races. In contrast, the validator with index $12$ achieves only a $7.9\%$ success rate, with performance decreasing monotonically as the validator index increases (Spearman correlation of $-1.0$).

We focus on the {\sf author} tiebreak and measure the ordering advantage of a low-numbered validator over high-numbered ones under all-pairs metric with no attack running, on Mysticeti and Bullshark protocols: on Mysticeti, the no-attack rate rises by $6.3$ points, while on Bullshark it is $49.98\%$, almost the same as base. This is because Mysticeti sorts blocks within a round by author identifier, so a low-identifier validator is placed ahead
of a high-identifier one nearly every round. Bullshark sorts by round only and leaves intra-round order to graph traversal, which carries no identity
signal.

We next attempt to fix Mysticeti's author tiebreak by replacing it with the digest of blocks (\cellref{P3.3}) and with a seed that combines the hash with a value drawn per commit (\cellref{P3.4}), both reducing the bias in measurement. Finally, we borrow the ordering idea of the time-based order-fairness protocols (e.g., Themis~\cite{kelkar2023themis}, DoD~\cite{nagda2026dag}, and FairDAG~\cite{kang2025fairdag}) by sorting same-round blocks on proposer timestamp with a digest fallback (\cellref{P3.5}).
A proposer writes its own timestamp, so it is worth asking whether it can simply lie. The rule puts the earliest stamp first, so winning a same-round race means claiming an early time. But the proposal-timestamp attack gains from the opposite, claiming a late one (\cellref{A2.6}). An attacker can pick an early stamp or a late one, not both, so buying priority inside a round costs it the advantage it was stamping for.
\end{proto}

\begin{proto}
\label{sec:find:leader}
\textbf{Leader rule.}
We run the silent-except-leader attack on Mysticeti, and change only how leaders are chosen.
Under the deployed rotation, which is round-robin, the attack lifts the rate by $19.8$ points (\cellref{P4.1}). This is because a fixed rotation is public, so an attacker knows in advance which rounds are its own and can plan to speak only in them. A random rotation removes that: the attacker cannot tell which rounds to save its blocks for.
Making the election stake-weighted and giving the attacker five times the average stake lowers the gain a little, to $17.4$ points (\cellref{P4.2}). This is because extra stake buys extra leader slots, and slots are worth more to a validator that publishes in all of them than to this attacker, which already publishes only in the rounds it anchors. The added slots therefore raise what the attacker would earn without attacking, buying it volume rather than position; Table~\ref{tab:cells} scores both stake settings against one common control, so that shows up as a smaller lift rather than a higher base. Stake purchases leadership, and this attack was never short of leadership.
Finally, under the leader-reputation rule (the leader scoring and schedule that Sui enables for Mysticeti in production), the rate falls by $1.4$ points (\cellref{P4.3}).
This is because the reputation derives from the recent involvement of each node in the protocol, and a node that becomes silent on most rounds can not hold the anchor position.
\end{proto}

\begin{proto}
\label{sec:find:tuning}
\textbf{Tuning.}
Tuning parameters shift the attack success rate too. In our experiments, AlephBFT's election seed swings the speculative success rate from $0.4$ to $96.0$, as the seed decides which hash rank wins the election. Similarly, on Mahi-Mahi, lengthening the wave from $\lambda{=}3$ to $\lambda{=}12$ under speculative takes the same-round rate from $51.98$ to $83.33$ (\cellref{P5.1}), because a longer wave leaves the attacker more rounds in which to search.
Since most of these parameters are specific to one protocol each, we report a single cell here and give the full set of sweeps, covering all six protocols, in \appref{sec:tuning}. The point is that every protocol carries parameters of its own that can shift an attack success rate, and which attack a given parameter shifts is set by the protocol's structure rather than by the parameter itself.
\end{proto}

\subsection{The Target Family}\label{secx:target}

\begin{target}
\label{sec:find:victims}
\textbf{Victims.}
We raise the number of simultaneous victims from one (\cellref{T1.1}) to three (\cellref{T1.2}), running fissure on Bullshark. Because the dimension is how many named targets the attacker aims at, we score it on the targeted committing-height metric. The single-victim row reuses the bribery run (\cellref{A5.2}); the three-victim row is a plain four-attacker campaign, so the two are each read against the same fair line rather than against one another.
With a {\sf single} victim, the success rate improves by $16.29$ points compared to the fair bullshark baseline (\cellref{T1.1}) due to the impact of bribery (as discussed in \cellref{A5.2}). However, targeting $3$ victims at the same time drops success against the victims by $11.9$ points.
This is because every validator (victim) an attacker blacklists is one fewer valid parent available to it, so refusing three leaves it choosing from a thinner set and degrades its own connectivity, while the pressure on any individual victim is reduced.
\end{target}

\begin{target}
\label{sec:find:metric}
\textbf{Metric.}
We take one set of committed orders and score it three ways: the all-pairs rate, the same-round rate, and the targeted committing-height rate. No experiment is re-run; only the definition of success changes. The order we score is the bribery run of A\ref{sec:find:bribe}, as bribery can demonstrate the impact of metrics more clearly.
Bribing three honest validators on Bullshark leads to $1.9\%$ reduction under all-pairs (\cellref{T2.1}), $10.1\%$ gain under same-round (\cellref{T2.2}), and $16.3\%$ gain under the targeted committing-height rate (\cellref{T2.3}).
The metrics ask different questions: all-pairs asks whether the attacker's blocks tend to precede the victim's anywhere in the order, targeted committing-height whether one particular attacking block beat one particular victim block. Fissure delays victims globally without necessarily winning any single same-round race, so the first answers yes and the second no. As a result, the reported success rates for ordering attacks are not comparable unless the metric and the control are both stated.
The fourth metric (\cellref{T2.4}), realized MEV, cannot be measured using the same run, as the bribery campaign carries no value model. We report realized MEV by running the silent-except-leader attack on Mysticeti under the uniform model instead. The run is shared with \cellref{T3.1} and the details are presented below.
\end{target}

\begin{target}
\label{sec:find:value}
\textbf{Value.}
All three rows run silent-except-leader on Mysticeti and differ only in the model. It changes nothing about the committed order, so it is the one dimension that moves only realized MEV. Each row is scored against a no-attack control drawn under the same model.
Under a uniform model, where every transaction is worth the same, the attack moves realized MEV from $58.0$ to $19.0$ (\cellref{T3.1}).
Silent-except-leader proposes only in the rounds the attacker leads and stays quiet in every other round, so it holds a better position in each race it enters but enters far fewer of them. As a comparison, all-pairs scores the fraction of races won and rises to $76.09$ (\cellref{A2.4}) while realized MEV
scores the value actually carried away, which needs blocks on the ledger, and those are what the attacker just declined to publish.
Under Pareto, where most of the value sits in a few transactions, the success rate is reduced by $73.7\%$ (\cellref{T3.2}) while under lognormal, whose tail is heavier still, its reduction is about $79.0\%$ (\cellref{T3.3}).
The value model basically keeps the direction of attack impact the same, so no value model reverses a gain to a loss, and the magnitude grows as the tail gets heavier, so the more the value concentrates in a few transactions, the more the choice of model matters.
\end{target}

\subsection{The Deployment Family}\label{sec:find:deploy}

\textbf{D 1--D 3.} Deployment covers how many validators there are (\dimref{D1}), how stake is spread across them (\dimref{D2}), and where they sit (\dimref{D3}). We vary each on its own and hold the attack fixed at silent-except-leader to show the impact more clearly.

Silent-except-leader gains $19.8$ points when the committee size $n{=}13$, gains $4.4$ at $n{=}25$ and gains $13.0$ at $n{=}49$ (\cellref{D1.1}--\cellref{D1.3}). Skewing stake toward the attacker leaves the gain near where equal stake puts it, $17.4$ against $19.7$ points (\cellref{D2.1}, \cellref{D2.2}). Spreading the committee over a wide area rather than one site does not move the rate: both arms reach the same $71.67\%$ (\cellref{D3.1}, \cellref{D3.2}), and giving one region a slower link than the others leaves it at $75.40\%$ (\cellref{D3.3}). 

Growing the committee from $n{=}13$ to $n{=}25$ on Mysticeti removes most of the gain, as the attacker competes against a denser frontier of honest blocks. It does not fall all the way, and most of the difference sits in the reference rather than the attack: between $n{=}25$ and $n{=}49$ the gain rises by $8.6$ points (\cellref{D1.2}, \cellref{D1.3}), but the attacked runs rise by only $3.7$ ($59.2$ to $62.9$) while the matched controls fall by $4.9$ ($54.8$ to $49.9$). Our attackers hold the lowest indices, so they begin with the free ordering advantage of P\ref{sec:find:tiebreak}, and that advantage thins as the committee grows: its concentration falls from $0.31$ at $n{=}13$ to $0.30$ at $n{=}25$ and $0.22$ at $n{=}49$. A larger committee leaves a low-index attacker less to inherit, so the control approaches $50\%$ and the same attack scores a larger gain against it.
Wide-area latency is likewise not a universal amplifier: it strengthens attacks built on the attacker's own delay and weakens attacks that depend on seeing other validators' blocks promptly. Neither a larger committee nor a geo-distributed deployment should be treated as a mitigation.

\section{Protocol-Specific Parameters (P\ref{dim:tuning})}\label{sec:tuning}

Tuning (P\ref{dim:tuning}) is the one dimension whose values are not shared across protocols, since each design exposes its own settings. Table~\ref{tab:cells} reports a single cell for it, so we summarize the full set of sweeps in Table~\ref{tab:tuning}; the raw runs behind them are in the artifact \cite{artifact}. Every sweep holds the protocol, committee size, and attacker budget fixed and changes one setting, and each entry below is the range of median ASR across the values swept, with $50\%$ as fair ordering. Sweeps are scored all-pairs, except where the setting governs an election, which is scored same-round instead; the Mahi-Mahi wave sweep below is one such case, and shares its endpoints with \cellref{P5.1}.

The sweeps come in two kinds. Most settings are \emph{protocol-native}: they exist in one design and have no counterpart elsewhere, such as a wave length or an election seed. Three are \emph{shared}, in that several implementations expose them under the same name: how many recent rounds stay hot in memory (cache depth $c$), how many rounds of old DAG state survive before pruning (garbage-collection depth $g$), and how long a node waits before retrying a failed synchronization ($t_{\mathrm{sync}}$). Bullshark exposes no native ordering parameter, so it is swept on the shared knobs alone.

\begin{table}[tbp]
\centering\scriptsize
\caption{Parameter sweeps across all six protocols, with raw runs in \cite{artifact}. Each cell is the range of median ASR over the values swept
($n{=}13$, $50\%$ neutral). A narrow range means the setting does not reach the attack.}
\label{tab:tuning}
\setlength{\tabcolsep}{3.5pt}
\renewcommand{\arraystretch}{0.92}
\begin{tabular}{@{}llccc@{}}
\toprule
\textbf{Protocol} & \textbf{Parameter} & \textbf{Fis.} & \textbf{Spec.} & \textbf{Slug.} \\
\midrule
\multicolumn{5}{@{}l}{\emph{Protocol-native settings}}\\
Narwhal-Tusk & workers $w{\in}\{1,2,4,8\}$        & $51.3$--$64.5$ & $58.3$--$62.1$ & $25.0$--$50.9$ \\
Mysticeti & leaders/wave $L{\in}\{2,4,6,8\}$      & $44.6$--$46.3$ & $45.4$--$49.0$ & $71.0$--$74.8$ \\
Mysticeti & wave $\lambda{\in}\{3,5,8,12\}$       & $43.8$--$48.4$ & $44.2$--$51.8$ & $66.0$--$73.6$ \\
Mahi-Mahi & leaders/wave $L{\in}\{2,4,6,8\}$      & $50.7$--$53.3$ & $54.2$--$65.0$ & $56.3$--$66.7$ \\
Mahi-Mahi & wave $\lambda{\in}\{3,5,8,12\}$       & $51.9$--$56.4$ & $\mathbf{52.0}$--$\mathbf{83.3}$ & $51.1$--$66.7$ \\
AlephBFT & lookahead $\kappa{\in}\{2,3,5,8\}$     & $52.6$--$52.8$ & $95.5$--$96.4$ & $\mathbf{48.8}$--$\mathbf{78.8}$ \\
AlephBFT & coord.\ delay $\{50,200,1000\}$\,ms    & $52.7$--$52.8$ & $95.7$--$95.8$ & $\mathbf{46.9}$--$\mathbf{74.7}$ \\
AlephBFT & hash seed $\{0,123,456\}$              & $46.5$--$52.8$ & $\mathbf{0.4}$--$\mathbf{96.0}$ & $49.3$--$58.3$ \\
Autobahn & slots $K{\in}\{1,2,4,8,16\}$           & $49.9$--$51.4$ & $49.9$--$50.1$ & $49.7$--$50.0$ \\
Autobahn & fast path $\{50\ldots1000\}$\,ms       & $49.9$--$50.0$ & $50.0$--$50.0$ & $49.8$--$50.0$ \\
\midrule
\multicolumn{5}{@{}l}{\emph{Shared core settings}}\\
Bullshark & cache $c{\in}\{1,2,5,20,50\}$         & $90.2$--$94.8$ & $87.4$--$93.8$ & $92.4$--$94.2$ \\
Bullshark & GC depth $g{\in}\{5,10,50\}$          & $91.6$--$93.4$ & $87.7$--$92.8$ & $92.9$--$94.1$ \\
Bullshark & $t_{\mathrm{sync}}\{0.5,2,5\}$\,s     & $90.6$--$92.8$ & $91.7$--$93.9$ & $92.8$--$94.3$ \\
Narwhal-Tusk & cache $c{\in}\{1,2,5,20\}$         & $\mathbf{33.3}$--$\mathbf{100.0}$ & $44.4$--$75.0$ & $55.0$--$75.0$ \\
Narwhal-Tusk & GC depth $g{\in}\{5,10,50\}$       & $\mathbf{66.7}$--$\mathbf{100.0}$ & $33.3$--$66.7$ & $52.9$--$80.0$ \\
Narwhal-Tusk & $t_{\mathrm{sync}}\{0.5,2,5\}$\,s  & $50.0$--$66.7$ & $58.3$--$66.7$ & $\mathbf{50.0}$--$\mathbf{100.0}$ \\
\bottomrule
\end{tabular}
\end{table}

Four patterns hold across the table, and all four follow the reading of \S\ref{sec:find}.

\paragraph{Tuning modulates an attack; it does not create one.} Autobahn stays within $1.4$ points of fair ordering on every setting and every attack, because a slot-based rule gives a proposer no shared seam to contest (\cellref{P2.3}); there is nothing for a parameter to open.
The same logic runs the other way on Bullshark, which never drops below $87\%$ on any shared setting: once certification hands an attacker the admission gate (P\ref{sec:find:cert}), no amount of cache, pruning, or timeout retuning takes it back.

\paragraph{Which attack a setting reaches is fixed by the structure.} On AlephBFT, the election lookahead and the coordination delay move sluggish by roughly $30$ and $28$ points while leaving fissure and speculative flat to within a point. Both settings change how long a validator may wait before its unit is counted, which is exactly the lever sluggish uses and neither of the other two does. A protocol's parameters are therefore not interchangeable defenses: each one reaches the attacks that consume the property it controls.

\paragraph{One setting dominates the rest.} AlephBFT's hash seed moves speculative from $0.4$ to $96.0$, a swing of $95.6$ points, which is the largest single effect anywhere in our measurements. The seed fixes which hash rank wins a round's election, so an attacker that knows it can grind toward that rank and one that does not is left searching (\cellref{A4.2}). A value chosen once by an operator, and documented as arbitrary, decides whether the election is grindable.

\paragraph{One protocol is genuinely tunable, and that cuts both ways.} Narwhal-Tusk is the only design whose shared settings move it across the whole range, with fissure running from $33.3$ to $100.0$ as cache depth changes. Its primaries seal headers from worker-produced batch digests, so these settings decide how long parent and batch information stays usable locally, and therefore what an attacker can still exclude before the next header is sealed. An operator can tune Narwhal-Tusk toward fairness, but the same width means a careless setting is as reachable as a careful one, and its readings vary more run to run than any other protocol here.

Narwhal-Tusk's header and batching parameters are swept in \cite{artifact} and not repeated here. The resource measurements reported there show that none of the shared settings shift throughput, latency, memory, CPU, or disk enough to change deployability, so on the protocols where retuning fails, it fails for free rather than at a cost worth trading.

\section{Mitigations and Their Limits}\label{sec:mitigation}

Every attack we measure works through one of three levers: what support a validator extends to another's block, which parents it references and which headers it signs; which locally available candidate a proposer picks; and when a block is sealed relative to its neighbors. Fissure and vote-withholding live on the first lever, speculative on the second, and sluggish and silent-except-leader on the third. All three sit \emph{between} blocks, not inside one, so a defense against them has to constrain \emph{inter-block} ordering. That is a different target from most existing fair-ordering work, which constrains \emph{intra-block} reordering within a single proposer's block~\cite{kelkar2020order,kelkar2023themis,cachin2022quick,nagda2026dag,kang2025fairdag}. We walk through four directions from that literature and ask, for each, which of our own attacks it would actually reach.

Two early defenses aim lower than manipulation and so do not reach any of our attacks. Censorship resistance~\cite{miller2016honey} only guarantees that correct transactions are eventually ordered, and reputation-based systems~\cite{asayag2018fair,kokoris2018omniledger,lev2020fairledger,crain2021red} only detect unfair censorship; neither stops a proposer from reordering the transactions it does include, so sandwiching passes through both untouched. \emph{Order-fairness}~\cite{nagda2024rashnu,kursawe2020wendy,kursawe2021wendy,kelkar2020order,kelkar2023themis,cachin2022quick} targets ordering manipulation directly, in three variants: hide content until commit, spread out who proposes, or order by time instead of by a validator-chosen key.

\paragraph{Hide content until commit.} Blind order-fairness~\cite{li2024sok} and content-agnostic ordering encrypt or secret-share transaction content~\cite{asayag2018fair,cachin2001secure,miller2016honey,stathakopoulou2021adding}, revealing it only once the order is fixed, so an attacker can neither recognize nor construct the block it wants to frontrun. Both leak through metadata and through client-leader collusion~\cite{kelkar2020order,kelkar2023themis,kursawe2020wendy}, and both add encryption and communication overhead. Hiding content would blunt speculative, which ranks candidates by a digest-sensitive score, but leaves fissure's reference-selection lever and sluggish's timing lever untouched, and it is not always applicable in the first place: Mahi-Mahi has no public mempool to hide, which is exactly why speculation there needs an outside price oracle rather than mempool access (\cellref{A4.3}).

\paragraph{Spread out who proposes.} Randomized leader or committee election~\cite{kiayias2017ouroboros,abraham2018solida,gilad2017algorand,lev2020fairledger,pass2017hybrid,asayag2018fair,yakira2021helix,malkhi2022maximal,spiegelman2022bullshark,keidar2021all,danezis2022narwhal} guarantees that many honest parties, not one, contribute to the final order, but an adversarial proposer can still order transactions unfairly inside its own turn, which is precisely the freedom silent-except-leader exploits by discarding every turn except the anchor ones. A more direct version of this idea replaces the ordering rule itself with a randomized one, e.g., combining on-chain randomness~\cite{gorman2025vraas} with block digests so ordering priority stops being predictable in advance. Of the four directions, this one comes closest to the lever we study, since it targets the same deterministic tiebreak we replace in \S\ref{sec:mitigation:fix}. Its costs are real: discarding the happen-before relationship among blocks can break data-dependent transactions, and randomization adds computation. Whether it survives contact with a live, concurrent DAG is open.

\paragraph{Order by time.} Time-based order-fairness~\cite{malkhi2022maximal} orders same-round transactions by when they were sent or received rather than by a validator-chosen key: client-side timestamps, measured propagation delay, or arrival order at each node. One instantiation has every node sign a timestamp per block it receives and orders by the median of the signed values; it holds only while honest nodes stay synchronized, breaks under a network-level delay attack, and adds a signature to every block. Client-side timestamps can simply lie, and measuring per-transaction network latency is hard under an asynchronous model with arbitrary delay, exactly the vulnerability Mysticeti's own default timestamp handling exposes, where an unbounded stamp buys a proposer $10.9$ points (\cellref{A2.6}). \S\ref{sec:mitigation:seeded} tests a bounded version of the same idea, ordering same-round blocks by proposer timestamp with a digest fallback, and finds that a self-referential design closes the hole: the stamp that would win the tiebreak is the same stamp that costs the attacker its commit placement, so gaming it in either direction gives nothing back.

\paragraph{Reorder after the fact.} A last option leaves the ordering rule alone and re-sorts the already-committed transactions afterward, typically by fee. This neutralizes any attack that works purely by manipulating block order, but reopens plain fee-based frontrunning: an attacker need only outbid the victim rather than fight over the DAG's structure. In a DAG, this trade is worse than in a single chain, because the attacker and victim need not even share a block, so there is no structural obstacle standing between an attacker and an outbid. \S\ref{sec:mitigation:gasprice} examines a deployed instance of exactly this mitigation, and the tie-breaking bug that lets it miss the common case.

\subsection{An MEV Mitigation on Mysticeti}

Mysticeti orders blocks inside a round by validator identifier, so with no attacker running a low-numbered validator takes the front position in $56.3\%$ of the races it shares with a high-numbered one, where $50\%$ would be fair\ifextend~\cite{mirzaei2026fair}\fi. Positions are handed out by identity rather than earned, which makes the tiebreak the natural target for a mitigation. Everything below applies only to protocols that break intra-round ties by author identity; Bullshark sorts by round alone and has no such bias to remove. We take four steps: the countermeasure Mysticeti already ships and why it misses, a one-line change that removes the bias for free, why the obvious version of that change can be cheated, and the version we recommend instead.

Countermeasures for biased ordering have been surveyed and grouped~\cite{yang2024sok}, and ours sits in the narrowest group: change the ordering rule and nothing else. The wider groups either enforce an ordering property directly, by batch-order fairness~\cite{kelkar2020order,kelkar2022order,kelkar2023themis}, a fair-ordering DAG construction~\cite{kang2025fairdag,putnik2026herring}, a leaderless commit rule~\cite{malkhi2022maximal}, or commit-before-see~\cite{wadhwa2024data}, or remove the information the attacker needs by encrypting transactions until the order is fixed~\cite{kavousi2025blindperm}. Those carry their own costs and impossibility limits; our aim is the cheapest change that removes the advantage we measure.

\subsection{The Fix Misses The Common Case}\label{sec:mitigation:gasprice}

Mysticeti already ships a mitigation aimed at exactly this bias, and it shows how narrow such fixes can be in practice. After consensus produces an order, the protocol re-sorts the committed transactions by gas price, highest first. Paying more buys priority, and equal-priority transactions are meant to be no longer decided by consensus position. However, the re-sort is a single call to the standard library's sort-by-key routine, and that routine is \emph{stable}: when two keys compare equal, it preserves their input order. The input order here is the consensus order, which is the very ordering the re-sort is meant to neutralize. So when two transactions offer the same gas price, the re-sort changes nothing and the low-numbered validator's transaction stays in front. The countermeasure engages only when one transaction pays \emph{strictly} more than another.

This is problematic because ties are the common case. Each validator quotes a reference gas price for the epoch, a low standard fee the network agrees to honor, and ordinary transactions simply pay that rate. Two transactions competing for the same opportunity therefore usually carry identical gas prices, and under those conditions the protocol falls back to precisely the identifier-based ordering the re-sort was added to remove. The lesson generalizes. A mitigation written as a re-sort inherits the tie-breaking behavior of whatever sort routine implements it. The choice between a stable and an unstable sort therefore decides, silently, whether the mitigation reaches the cases that matter most. Removing the bias requires changing the ordering rule itself.

\subsection{A minimal, Free Fix}\label{sec:mitigation:fix}

The advantage exists because the ordering rule breaks intra-round ties by author index. That tiebreak only needs to be \emph{deterministic and consistent across validators}; any such rule preserves safety, and the reference implementation's own comment says as much. Replacing the author index with the block \emph{digest}, sorting within a round by $H(\text{block})$ instead of by author, is such a rule. We call it \textsc{FixTiebreak}.

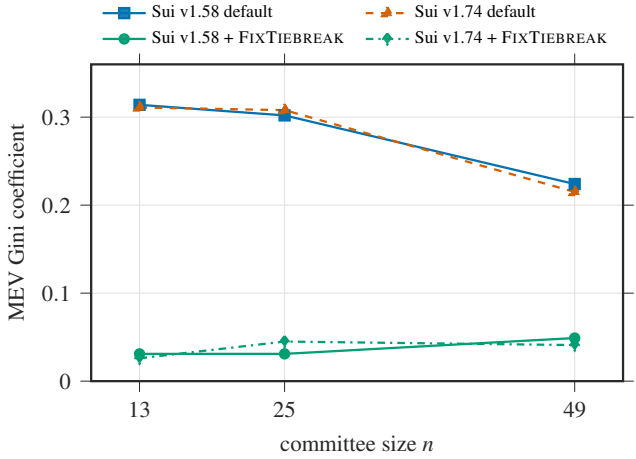
\begin{figure}[!t]
\centering
\resizebox{\columnwidth}{!}{
\begin{tikzpicture}
\begin{axis}[
  mevaxis, width=9.0cm, height=6.0cm,
  xlabel={committee size $n$}, ylabel={MEV Gini coefficient},
  xmin=9, xmax=53, ymin=0, ymax=0.36,
  xtick={13,25,49}, ytick={0,0.1,0.2,0.3},
  legend style={
    at={(0.5,1.02)}, anchor=south, legend columns=2, draw=none, fill=none,
    font=\scriptsize, /tikz/every even column/.append style={column sep=6pt},
  },
  legend cell align=left,
]
\addplot[cBull, solid, mark=square*, mark size=2pt] coordinates {(13,0.314)(25,0.302)(49,0.224)};
\addlegendentry{\Sui{} v1.58 default}
\addplot[cSui, dashed, mark=triangle*, mark size=2.4pt] coordinates {(13,0.311)(25,0.308)(49,0.215)};
\addlegendentry{\Sui{} v1.74 default}
\addplot[cFix, solid, mark=*, mark size=1.8pt] coordinates {(13,0.031)(25,0.031)(49,0.049)};
\addlegendentry{\Sui{} v1.58 + \FixTB}
\addplot[cFix, dash dot, mark=diamond*, mark size=2.2pt] coordinates {(13,0.026)(25,0.045)(49,0.041)};
\addlegendentry{\Sui{} v1.74 + \FixTB}

\end{axis}
\end{tikzpicture}}
\caption{Concentration of the same advantage as the committee grows, measured as a Gini
coefficient over per-validator MEV. It stays near $0.31$ at $n{=}13$ and $n{=}25$ and eases to
about $0.22$ at $n{=}49$, so a larger committee dilutes the advantage without removing it, while
\textsc{FixTiebreak} holds it near $0.03$ at every size.}
\label{fig:scaling}
\end{figure}

On Mysticeti it moves the no-attack rate from $56.30\%$ to $54.29\%$, a fall of $2.0$ points (\cellref{P3.3}), and Figures~\ref{fig:indextax} and~\ref{fig:scaling} show the rest of the effect: the win-rate staircase collapses to a flat $50\%$ band, and the Gini falls from $0.31$ to $0.03$ at every committee size. It costs nothing.

\begin{figure}[!t]
\centering
\resizebox{\columnwidth}{!}{
\begin{tikzpicture}
\begin{axis}[
  mevaxis, width=9.6cm, height=6.6cm,
  xlabel={validator index (ID)}, ylabel={same-round win-rate (\%)},
  xmin=-0.5, xmax=12.5, ymin=0, ymax=100,
  xtick={0,2,4,6,8,10,12}, ytick={0,25,50,75,100},
  legend style={
    at={(0.5,1.02)}, anchor=south, legend columns=2, draw=none, fill=none,
    font=\scriptsize, /tikz/every even column/.append style={column sep=6pt},
  },
  legend cell align=left,
]
\addplot[cbGray, dashed, line width=0.8pt, forget plot] coordinates {(-0.5,50) (12.5,50)};
\node[cbGray, font=\scriptsize, anchor=south west] at (axis cs:-0.4,50.5) {fair (50\%)};

\addplot[cBull, solid, mark=square*, mark size=1.6pt] coordinates {
 (0,92.0)(1,85.2)(2,78.3)(3,71.5)(4,63.1)(5,55.6)(6,48.9)(7,41.7)(8,35.6)(9,28.6)(10,21.6)(11,14.9)(12,7.4)};
\addlegendentry{\Sui{} v1.58 default}
\addplot[cSui, dashed, mark=triangle*, mark size=2.0pt] coordinates {
 (0,91.1)(1,85.0)(2,78.2)(3,71.7)(4,62.9)(5,55.4)(6,48.5)(7,42.0)(8,36.0)(9,28.7)(10,21.7)(11,15.2)(12,7.9)};
\addlegendentry{\Sui{} v1.74 default}
\addplot[cFix, solid, mark=*, mark size=1.5pt] coordinates {
 (0,49.3)(1,54.0)(2,50.4)(3,51.1)(4,49.0)(5,54.1)(6,45.4)(7,49.6)(8,48.7)(9,50.4)(10,52.3)(11,49.5)(12,46.3)};
\addlegendentry{\Sui{} v1.58 + \FixTB}
\addplot[cFix, dash dot, mark=diamond*, mark size=2.0pt] coordinates {
 (0,47.6)(1,48.4)(2,51.3)(3,53.3)(4,49.9)(5,53.4)(6,47.0)(7,48.9)(8,51.8)(9,51.8)(10,49.8)(11,50.6)(12,46.2)};
\addlegendentry{\Sui{} v1.74 + \FixTB}

\node[align=left, font=\scriptsize, anchor=south west, fill=white, fill opacity=0.85,
      text opacity=1, rounded corners=1pt, draw=black!15]
  at (axis cs:-0.2,2) {default: Spearman $=-1.00$, Gini $=0.31$\\ \FixTB: Spearman $\approx0$, Gini $=0.03$};
\end{axis}
\end{tikzpicture}}
\caption{Same-round win rate against validator index, with no attacker running. The two shipped
builds sort by $(\text{round},\text{author})$ and trace the same staircase, from $91.1\%$ at
index $0$ to $7.9\%$ at index $12$; \textsc{FixTiebreak} flattens both to the fair line.}
\label{fig:indextax}
\end{figure}
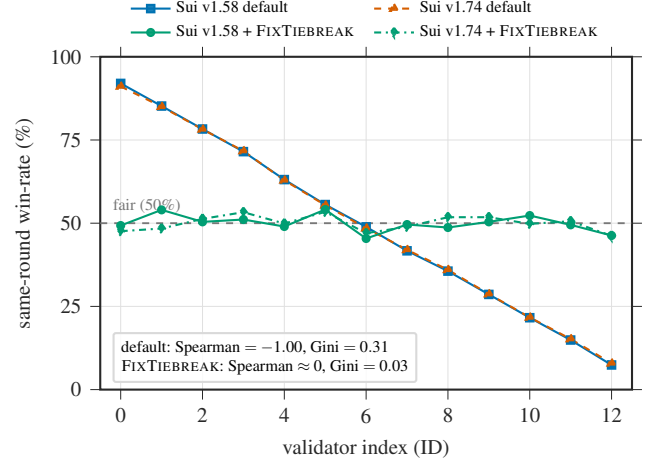

\begin{figure}[t]
\centering
\resizebox{\columnwidth}{!}{\begin{tikzpicture}
\begin{axis}[
  mevaxis, width=9.2cm, height=6.2cm,
  xlabel={grinding attempts $G$ (block digests tried)}, ylabel={grinder win-rate (\%)},
  xmode=log, log basis x=2,
  xmin=0.9, xmax=1200, ymin=45, ymax=102,
  xtick={1,4,16,64,256,1024}, xticklabels={1,4,16,64,256,1024},
  ytick={50,60,70,80,90,100},
  legend pos=south east,
]
\addplot[cbGray, dashed, line width=0.8pt, forget plot] coordinates {(0.9,50)(1200,50)};
\addplot[cbInk, solid, line width=1.0pt, no marks, domain=1:1024, samples=80]
  {100*x/(x+1)};
\addlegendentry{analytic $G/(G{+}1)$}
\addplot[cAttack, only marks, mark=*, mark size=2.2pt, patDefault] coordinates {
 (1,49.9)(2,66.7)(4,80.0)(8,88.7)(16,94.1)(32,97.0)(64,98.5)(128,99.2)(256,99.6)(1024,99.9)};
\addlegendentry{Monte-Carlo (40k rounds)}
\draw[cbInk, ->, >=stealth, line width=0.6pt] (axis cs:64,86) -- (axis cs:64,97.2);
\node[cbInk, font=\scriptsize, anchor=north, align=center] at (axis cs:64,85.5)
  {$G{=}64$ (\textmu s of work)\\ $\to 98.5\%$};
\node[cbGray, font=\scriptsize, anchor=south west] at (axis cs:1.1,50.6) {fair (50\%)};
\end{axis}
\end{tikzpicture}}
\caption{The naive digest tiebreak is grindable. Trying $G$ candidate digests and keeping the smallest wins a fraction $G/(G{+}1)$ of intra-round races; Monte-Carlo matches the analytic to $0.1\%$. Sixty-four hashes ($\mu$s of work) already buy a $98.5\%$ win-rate.}
\label{fig:grind}
\end{figure}
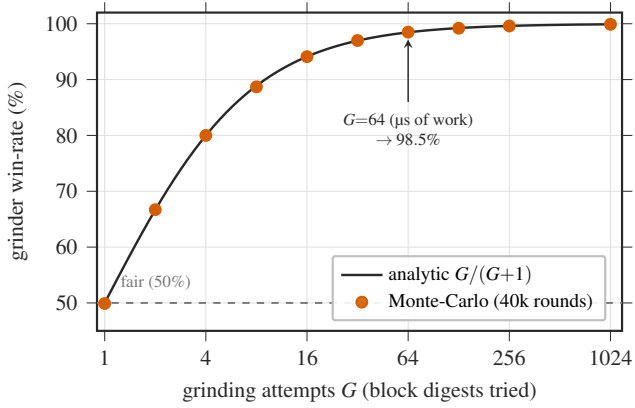

The number of committed blocks is unchanged, because the change decides only which deterministic key orders a commit, not whether the commit forms. The digest tiebreak is a one-line, safety-preserving, performance-neutral fix.

The fix is complete for the bias it targets and silent about a second one. Separating the two takes two of our metrics. Same-round scores only block pairs committed in the same round; all-pairs scores every attacker/victim pair across all rounds. With no attacker running, the same-round rate runs from $91.1\%$ at the lowest index down to $7.9\%$ at the highest; once the digest tiebreak is armed, every index sits within four points of $50\%$. The intra-round advantage is gone exactly as intended. The all-pairs rate over the same runs moves only from $56.3\%$ to $53.7\%$. Most attacker/victim pairs are drawn from \emph{different} rounds, and an intra-round tiebreak cannot touch those. A residual advantage therefore survives the fix, and it originates elsewhere in the protocol.

\subsection{Changing the Rule Moves the Floor}\label{sec:mitigation:orderrule}

The ordering rule is a knob inside a single implementation, so we can hold the protocol fixed and vary only that rule, each arm against a baseline produced under the same rule (Table~\ref{tab:orderrule}). Removing identity from the rule steadily lowers what the protocol gives away for free, from $56.3\%$ to $53.7\%$ to $52.7\%$. The plain round-only rule that leaves intra-round order to graph traversal is the fairest of the three, slightly ahead of the digest tiebreak. But none of the three changes what either attack achieves. Fissure stays inert under all of them, and silent-except-leader scores within a third of a point of itself throughout ($75.9$, $75.9$, $75.6$).

This can mislead a defender, so it is worth stating plainly. Silence gains $19.6$ points over the production rule but $23.0$ over round-only, purely because the reference it is measured against has fallen. Read as raw rates, the fix appears to have backfired. It has not. It removed a free advantage that was inflating the baseline, and the attack was always worth what it is now seen to be worth.

Silent-except-leader exploits what blocks reach anchor positions, not how ties inside a round are broken, so a tiebreak change leaves it untouched. Around $2.7$ points of structural advantage survive even under round-only. That residual is cross-round, and no intra-round rule can reach it. \textsc{FixTiebreak} is a defense against a structural bias, not against an adversary.

\begin{table}[tbp]
\centering\small
\caption{Ordering rule varied inside one protocol, with a matched no-attack baseline under each rule. All-pairs ASR, $n{=}13$, medians of five runs. Rules are written with $r$ the block's round and $\iota$ its proposer's index. Removing identity from the rule lowers the baseline but leaves both attacks where they were. The rules are compared within this campaign; in the attack-space campaign of Table~\ref{tab:cells} the digest rule reads $54.29$ with no attacker and $75.62$ under silence, a run-to-run spread that does not change the ranking.}
\label{tab:orderrule}
\footnotesize
\setlength{\tabcolsep}{3.5pt}
\begin{tabular}{@{}l r r r@{}}
\toprule
\textbf{Ordering rule} & \textbf{No atk.} & \textbf{Fissure} & \textbf{Silence} \\
\midrule
$(r,\iota)$, production & $56.30$ & $56.51$ & $75.94$ \\
$(r,\text{digest})$, \textsc{FixTiebreak} & $53.73$ & $54.78$ & $75.94$ \\
round only, traversal & $\mathbf{52.66}$ & $53.50$ & $75.62$ \\
\midrule
\emph{gain over own base.} & --- & $+0.2$--$1.1$ & $+19.6$--$23.0$ \\
\bottomrule
\end{tabular}
\end{table}

\subsection{The Naive Fix is Grindable}\label{sec:mitigation:grind}

The digest tiebreak replaces a bias a validator cannot cheat with one it can. A proposer chooses its own block's content, and therefore its digest. It can assemble $G$ candidate blocks, varying a nonce or the transaction set, and broadcast the one whose digest sorts earliest. Sorting by digest is sorting by a uniform key, so the best of $G$ independent draws lands, in expectation, in the $1/(G{+}1)$ quantile: the grinder wins a fraction $G/(G{+}1)$ of its intra-round races.

Figure~\ref{fig:grind} plots that formula against a Monte-Carlo draw of the same experiment, which tracks it to within $0.1\%$ over four orders of magnitude. At $G{=}1$, meaning no grinding, the grinder sits at the fair $50\%$. By $G{=}64$ (sixty-four hashes, microseconds of work per block) it wins $98.5\%$ of its races, and $G{=}1024$ reaches $99.9\%$. We report this as a property of the sort key rather than as a protocol measurement, because the arithmetic follows from the key alone.

The digest tiebreak converts a \emph{deterministic} unfairness keyed on a fixed identity into a \emph{grindable} one keyed on spare compute. The author index is something a validator cannot change; the digest is something anyone can search. The author tiebreak at least advantages a fixed, publicly known set of validators, whereas the grindable tiebreak advantages whoever spends the most hashes, on every block.

\subsection{A grind-resistant tiebreak}\label{sec:mitigation:seeded}

The tiebreak key is chosen by, and known to, the proposer at the moment the block is built. Four rules sit at the corners of a small design space, and the one we recommend is the corner that is fair \emph{and} out of the proposer's control:

\begin{itemize}[leftmargin=1.4em,itemsep=1pt]
\item $(\text{round},\ \text{author})$, deterministic and unfair. Every validator's position relative to every other is fixed by identity; the low index is permanently in front. It is \emph{not} grindable, because a validator cannot cheaply change its identity.
\item $(\text{round},\ \text{digest})$, grindable. Fair in expectation over an honest proposer's block, but a proposer controls its own digest and can grind it, so the rule stops being fair the moment anyone bothers to.

\item Round only, with intra-round order left to graph traversal. This measured the \emph{fairest} of the three rules we ran ($52.7\%$, against $53.7\%$ for the digest rule and $56.3\%$ for production), and it is not grindable through the block digest, because the digest plays no part in it. We stop short of recommending it because we did not establish that traversal order is beyond a proposer's influence, and an ordering rule is only as good as its worst manipulable input. Establishing that is worthwhile future work.

\item $(\text{round},\ H(\text{digest}\parallel s))$ with $s$ a per-commit value \emph{unpredictable at block-creation time}, for instance derived from the committed leader's digest or from a consensus randomness beacon. This is fair \emph{and} grind-proof. Because $s$ is not known when the block is built, grinding the digest yields no control over the final key, so the intra-round order becomes a lottery no one can steer.
\end{itemize}

That such a seed is not already needed for leader election is instructive. Leader election in Mysticeti is seeded on the round number alone, with no proposer-controlled input, and is consequently \emph{not} grindable: a validator cannot buy extra leader slots by varying block content. The designers starved the leader seed of adversarial entropy but left the intra-round tiebreak keyed on identity. The rule we recommend extends the leader-election discipline to the tiebreak.

That condition is easy to state and easy to lose, so it is worth showing what a \emph{predictable} seed buys. Measured as a passive bias, a seeded rule with a fixed, published seed behaves like the digest rule and no better: on Mysticeti it moves the no-attack rate from $56.30\%$ to $54.10\%$, a fall of $2.2$ points (\cellref{P3.4}), against $54.29\%$ for the digest. Measured against an attacker, it collapses. AlephBFT's election extension applies exactly this construction, sorting candidates by hash and rotating the order by a configured seed, and its own source describes the rotation as breaking ``the attacker's ability to grind a globally best hash''. It does so only against an attacker who grinds for the \emph{smallest} hash. The rotation fixes the winning rank at $\mathrm{seed} \bmod \ell$, with $\ell$ the round's candidate count, so an attacker who knows the seed knows which rank to aim for. Aiming at the corresponding quantile of the hash space instead of at the minimum takes the same configuration from $0.3\%$ to $77.8\%$ (\cellref{A4.2}), using no privileged information. A seed fixed at configuration time is therefore not a weaker version of the mitigation but a different object: the rule is grind-proof only while the seed is unknown \emph{at block-creation time}.

Ordering same-round blocks by proposer timestamp, with the digest as a tiebreak between equal stamps, lowers Mysticeti's no-attack rate further, from $56.30\%$ to $52.50\%$, a fall of $3.8$ points (\cellref{P3.5}), below the author, digest, and seeded rules alike. It also resists manipulation of its own sort key. An attacker that stamps its blocks earlier gains nothing ($-0.5$ points), and one that stamps them later drives itself to exactly the fair $50.00\%$ line, because the stamp that wins the tiebreak is the stamp that damages its commit placement. We report this as the ordering \emph{rule} of the order-fair family rather than as an implementation of its fairness guarantee.

\subsection{The Limits of The Whole Family}\label{sec:mitigation:limits}

Every rule discussed here decides the order \emph{within} a round, and our three-rule comparison showed that none of them changes what either attack achieves. They are worth deploying, because the advantage they remove is real, permanent, and free to whoever holds a low identifier. They are not a mitigation against an adversary. Defending against silent-except-leader means changing what a leader round is worth, which is a change to the ordering rule (P2) rather than to the tiebreak, and we leave it open.

Every number in this paper is a difference between an attack run and a baseline taken at the same attacker and victim placement, so the identifier-based advantage sits in both and subtracts out. That is cancellation by differencing, not a decomposition, and the \textsc{FixTiebreak} runs let us check how exact it is by measuring one attack with the advantage present and then removed. Under the author tiebreak silent-except-leader gains $19.79$ points ($56.30$ to $76.09$); with the digest tiebreak armed it gains $21.33$ ($54.29$ to $75.62$). Exact cancellation would make these agree, and they differ by $1.54$ points, or $1.34$ if the author arm is read against the dedicated no-attack control at $56.10\%$.

The residual is the interaction between the attack and the bias: staying silent changes which rounds the attacker occupies, hence how many attacker/victim pairs fall in the same round, and the tiebreak reaches only those pairs. Against effects of $20$ to $46$ points, it changes no conclusion here, but our differences should be read as close approximations of an attack's marginal contribution rather than exact separations.

\section{Discussion and Limitations}\label{sec:disc}

This section gives a detailed discussion of our experimental results and possible limitations of the study.

\paragraph{What the map reveals.}
Read across the cells (Figure~\ref{fig:designspace}), the attack space delivers one consistent message: what an attacker can do to a DAG-BFT protocol is decided by that protocol's own construction, not by how hard the attacker tries. The two constructions we measure first-hand are exposed through different doors and closed on each other's. Certification creates an admission gate, which is what makes fissure against a victim worth $42.9$ points; without that gate the identical attack is worth $4.3$ (\cellref{P1.2}). A leader-anchored linearization makes leader rounds positionally valuable, which is what makes staying silent worth $19.8$ points (\cellref{P2.1}); without it the identical attack is worth $-0.7$. The same asymmetry appears in the six-system breadth cell (Fig.~\ref{fig:primitive}), where fissure succeeds on the certified systems and speculative succeeds where the protocol gives a proposer freedom over candidate blocks. Changing the \emph{action} points the same way. A sandwich needs the attacker on both sides of its victim, and Bullshark takes the triplet-sandwich rate from $50.00\%$ to $81.82\%$ (\cellref{A1.3}), because the front side of that bracket is precisely what the admission gate hands over. One structural property therefore explains two results that look unrelated: why fissure works at all, and why sandwiching against Bullshark succeeds through the same gate.

The practical implication for a designer is narrow and useful. The question is not whether a protocol is vulnerable to MEV in general, but which of two structural properties it has, because each one names the attacks that follow from it and the attacks that do not.

\paragraph{The economics follow the structure.}
Because the two families differ in whether they need participants, so do their costs. The attacks that target another validator's position, fissure among them, need a coalition: coordination is worth up to $18.25$ points (\cellref{A3.3}), and a budget substitutes directly for compromise, with three bribed honest validators outperforming four genuine Byzantine ones (\cellref{A5.2}, \cellref{A6.2}). Silent-except-leader repositions only the attacker's own blocks, so it is fully effective at a single validator, gains almost nothing from adding independent attackers ($75.00\%$ solo versus $76.09\%$ competing, \cellref{A3.1}, \cellref{A3.2}), and an identical bribe buys exactly zero. An adversary's cheapest strategy is thus a property of the target rather than of the attack (A\ref{sec:find:collude} and A\ref{sec:find:bribe}), which is not a distinction the prior literature draws.

\paragraph{Negative results bound the space.}
Systematization is credible only if it records where attacks fail, and several of our probes returned clean negatives. Mysticeti's default leader election cannot be ground: its schedule derives from the round number alone, so no amount of block-content search buys additional leadership (\S\ref{sec:mitigation:seeded}). AlephBFT's election, seeded on a configured constant rather than on the round, is grindable, and that contrast is what identifies the round-number schedule as the discipline worth copying. The leader-reputation mechanism resists targeted abuse better than its exclusion-based design suggests: a colluding minority can measurably lower a rival's score but could not reliably push a \emph{specific} target below the demotion threshold, because round-to-round variance dominates the small signal a sub-threshold coalition injects, and the attack sometimes demotes a colluder instead. Attacking three victims at once extracts less than attacking one (T\ref{sec:find:victims}). And silent-except-leader destroys itself if pursued too far: once enough validators fall silent that the remaining proposers cannot form the quorum needed to advance a round, the protocol falls back to timeout-driven progress and the attacker's advantage disappears.

\paragraph{Generativity: a predicted cell, refuted.}
An attack space earns its keep when it turns intuition into falsifiable predictions. The Deployment$\times$Adversary interaction suggests one: skewing stake toward a silent attacker should deepen its advantage, since a richer validator ought to dominate. We tested this empty cell and the prediction fails (\S\ref{sec:find:deploy}). Making the election stake-weighted and giving a lone silent-except-leader attacker five times the average stake left it slightly \emph{worse} off than equal stake, and both codebases agree on the direction. The map explains why: stake buys leader slots, and this attack was never short of leader slots, since it already publishes only in the rounds it anchors. The extra slots raise the no-attack rate the attack is measured against without buying it any better position, so concentrating stake shrinks the lift instead of deepening it.

\paragraph{Threats to validity.}
\ifextend
The verification discipline of \S\ref{sec:method} is not precautionary: both failure modes it guards against bit us. Requiring evidence that a mechanism fired changed one of our own results by an order of magnitude, since an early measurement put coordination at $2$ points because the control was already colluding.
\fi
 A third trap emerged while we were checking the second. Evidence that a hook fired is written to the log, so obtaining it means raising the log level, and the log level is not inert: holding everything else fixed and changing only that setting moved a no-attack same-round rate from $94.2\%$ to $87.4\%$ over five repetitions each, with non-overlapping ranges. An attack and its control must therefore be measured at the same verbosity.

\paragraph{Limitations.}
Our measurements come from real multi-node runs in a controlled testbed rather than a public mainnet, at committee sizes up to $n{=}49$, so absolute magnitudes on a live network may differ even though the mechanisms, being consequences of the ordering rule, should not. We quantify MEV with a positional rate and a value-weighted proxy rather than settled dollars, grounding position in execution as \S\ref{sec:space} describes, and our value distributions are synthetic, used to test whether the ordering conclusions survive a change in the value model rather than to predict revenue. Two protocol dimensions cannot be varied like the others: a tiebreak or a tuning parameter is a setting inside one implementation, whereas whether blocks are certified and whether the ordering rule anchors on a leader are what a protocol \emph{is}, so P1 and P2 rest on comparison across implementations rather than a switch thrown inside one. We reduce that risk by measuring two protocols on two independent codebases each, which agree to within $1.1$ and $2.2$ points, but those two dimensions should be read as comparative and the rest as controlled. We measure all four actions in A1, yet the controlled single-dimension cells elsewhere all use frontrunning, so how the action interacts with the other dimensions rests on four protocols rather than throughout. Censorship is the one action no positional metric can score, since a censored block has no position, so \cellref{A1.4} is read on inclusion. Finally, the cells spanning all six systems run one configuration each, so what they support is breadth rather than mechanism.

\paragraph{Generality.}
The structural results in this paper follow from three protocol choices: the admission rule (P\ref{dim:type}), the ordering rule (P\ref{dim:ordering}), and the tiebreak (P\ref{dim:tiebreak}), so they transfer to any system sharing those choices rather than to a particular codebase. Concretely, a protocol that certifies blocks before admitting them to the DAG inherits the exposure to fissure; a protocol whose ordering rule anchors on a leader inherits the exposure to silent-except-leader; and a protocol that breaks intra-round ties by author identity inherits a fixed assignment of ordering positions to its validators by identifier, with no attacker present. These are independent choices, and the tiebreak in particular does not follow from the DAG type: Narwhal-Tusk and Bullshark are both certified yet break ties by digest and by round respectively, so neither carries an identity bias, while Mysticeti breaks them by author and carries a large one. A designer can consult the three rules in their own protocol and read off which results apply.

\section{Conclusion}\label{sec:conc}

This paper organized MEV attacks on DAG-BFT protocols, previously studied one system at a time under different definitions of success, into a single attack space, and measured one experiment per value across six production protocols. The results show that protocol structure, not attacker effort, decides what an adversary can do (\S\ref{secx:protocol}). In particular, a certified DAG's admission gate lets an attacker deny a victim inclusion; an uncertified, leader-anchored DAG instead rewards a lone attacker who simply stays silent outside its own rounds. The same split governs cost: attacks that reposition another validator's blocks need a coalition, and a budget substitutes for compromise, while attacks that reposition only the attacker's own blocks need neither (\S\ref{secx:adversary}).
Two further findings matter. First, the metric is itself part of the attack space: scoring one committed order in three defensible ways moved the success rate substantially and reversed which countermeasures looked effective (\S\ref{secx:target}). Second, intuition is an unreliable guide to mitigation: skewing stake toward a silent attacker was predicted to help it and instead left it slightly worse off (\S\ref{sec:find:deploy}), and removing identity from the intra-round tiebreak closes only part of the bias it creates while leaving other attacks untouched. Defending a DAG-BFT protocol starts with identifying which structural gate it actually has, since that gate, not the attacker's resources, is what a mitigation must close.
\section*{Ethical Considerations}
\label{sec:ethics}

This paper studies attacks on deployed consensus protocols, so we set out the stakeholders, the
harms we could plausibly cause, and what we did about them.

\noindent\textbf{No live system was attacked.} Every measurement in this paper comes from private
deployments of open-source protocol implementations, running on our own machines in a
resource-capped testbed. We never sent a transaction to a public network, never interacted with
a live validator set, never touched third-party funds, and never observed or handled any real
user's transaction. The only data we collect is the committed block order produced by validators
we ourselves operate. There are no human subjects and no personal data, so no institutional
review was applicable.

\noindent\textbf{We add no capability an adversary does not already have.} Every attacker behavior we
implement is a choice an honest validator is already permitted to make under the protocol:
which valid parents to reference, when to propose, and what to put in its own block
(\S\ref{sec:model}). We do not weaken signature checking, quorum-intersection checks or validity
predicates, and a continuous-integration guardrail rejects any change that touches that code
(\S\ref{sec:method}). Our artifact therefore does not hand anyone a capability they lacked; it
automates and measures choices that are already available to every validator in a committee, and
it cannot be pointed at a network the operator does not already control.

\noindent\textbf{The most serious finding is not an exploit.} The largest effect we report requires no
attacker at all. It follows from an ordering rule in deployed code that settles ties by validator
identifier, and it is visible to anyone who reads that code (P\ref{sec:find:tiebreak}). We are
describing a property of a published system rather than a technique we invented, and the same is
true of the gas-price countermeasure whose stable sort fails to fire on ties
(\S\ref{sec:mitigation:gasprice}).

\noindent\textbf{Disclosure.} Three findings concern systems in production or in public use: the
identifier-based tiebreak and the gas-price re-sort in the uncertified system we measure, and
the seeded-rotation mitigation in an election-based implementation, which we show is defeated by
public configuration alone (A\ref{sec:find:info}). We reported the first two to the vendor's
published security address on 25 August 2026, before submitting this work anywhere, and with the fix
attached rather than the finding alone: a one-line change to the tiebreak, the condition under
which a seeded rule remains sound, and the corresponding change to the equal-fee case of the
gas-price re-sort. We offered to hold publication for a coordinated disclosure date.
\emph{No response had been received at the time of submission.}

The third finding we did not report, and we state the reason rather than leave it to be inferred.
That project publishes no security contact: its bug-bounty programme is no longer live, its
repository carries no security policy, private vulnerability reporting is disabled, and the
codebase has not been updated in over a year. We judged a general-purpose contact address an
inadequate channel for a report we could not confirm would reach anyone, and the finding does not
warrant the alternative of publishing it to an open issue tracker. Two things bound the residual
risk. None of the three is a memory-safety or key-compromise bug that a reader could turn into an
immediate incident; each is a property of an ordering rule that its own source code already
states. And none confers an ability a validator lacks: the actions are
the ones any committee member may already take, which is the whole point of the threat model in
\S\ref{sec:model}.

\section*{Open Science}
\label{sec:openscience}

We intend every number in this paper to be reproducible by a third party, and the measurement
discipline the paper argues for is only checkable if the artifacts are available.

\noindent\textbf{What we release.} Three things. First, the six instrumented protocol
implementations, each carrying its attacker behaviors as environment-gated hooks: with the variable unset a hook returns before the default path, so the compiled binary is byte-identical
to upstream and honest validators always run the disarmed build. The repository lists every hook
with the variable that arms it. Second, the drivers and scorers, including the implementation of
every metric in \dimref{T2} and the reimplementation of the prior targeted metric that we
verified against its reference on identical logs. Third, the per-cell measurements behind
Table~\ref{tab:cells}: one row per repetition rather than a median, with each attack arm beside
the control arm it is scored against, so a reader can recompute a lift rather than take it on
trust.

\noindent\textbf{What a reproducer needs to know.} Two of our findings are about measurement rather
than about attacks, and both bear on reproduction. A cell is only meaningful against its own
matched control, so the artifact ships attack and control as a pair rather than as separate
runs. And the log verbosity that reveals a mechanism marker also perturbs the timing the
committed order depends on (\S\ref{sec:disc}), so the harness pins one verbosity across both arms
of a comparison and records it alongside the result. A reproducer who changes that setting should
expect a different baseline, and we give ours.

\noindent\textbf{Availability.} The artifact~\cite{artifact} is anonymised for review. It contains the six
instrumented implementations, the campaign drivers and scorers, the per-cell CSVs behind
Table~\ref{tab:cells}, and the figure pipeline, together with a reference listing every hook
and every ordering rule the mitigation evaluation switches between. The raw committed-order logs
the cells were scored from run to roughly half a gigabyte and are archived separately rather
than shipped in the repository; the scorers that turn those logs into the released CSVs are
included, so the pipeline is reproducible end to end from archived logs. On acceptance we
replace this link with a non-anonymous, archived version carrying a stable DOI and the commit
each measurement was taken at.

\balance

\bibliographystyle{plainurl}
\bibliography{_blockchain,_privacy,_system}

\begin{thebibliography}{10}

\bibitem{artifact}
{Artifact: Competition, Collusion, and Corruption --- code, drivers, scorers and per-cell results}.
\newblock \url{https://anonymous.4open.science/r/MEV-C729}, 2026.

\bibitem{abraham2018solida}
Ittai Abraham, Dahlia Malkhi, Kartik Nayak, Ling Ren, and Alexander Spiegelman.
\newblock Solida: A blockchain protocol based on reconfigurable byzantine consensus.
\newblock In {\em Int. Conf. on Principles of Distributed Systems (OPODIS)}. Schloss Dagstuhl-Leibniz-Zentrum fuer Informatik, 2017.

\bibitem{aptos}
{Aptos}.
\newblock Aptos: The foundation for a new digital economy.
\newblock \url{https://aptosnetwork.com/}, 2026.
\newblock Accessed 2026.

\bibitem{arun2025shoal++}
Balaji Arun, Zekun Li, Florian Suri-Payer, Sourav Das, and Alexander Spiegelman.
\newblock Shoal++: High throughput dag bft can be fast and robust!
\newblock In {\em Symposium on Networked Systems Design and Implementation (NSDI)}. USENIX Association, 2025.

\bibitem{asayag2018fair}
Avi Asayag, Gad Cohen, Ido Grayevsky, Maya Leshkowitz, Ori Rottenstreich, Ronen Tamari, and David Yakira.
\newblock A fair consensus protocol for transaction ordering.
\newblock In {\em Int. Conf. on Network Protocols (ICNP)}, pages 55--65. IEEE, 2018.

\bibitem{babel2025mysticeti}
Kushal Babel, Andrey Chursin, George Danezis, Anastasios Kichidis, Lefteris Kokoris-Kogias, Arun Koshy, Alberto Sonnino, and Mingwei Tian.
\newblock Mysticeti: Reaching the latency limits with uncertified dags.
\newblock In {\em Network and Distributed Systems Security Symposium (NDSS)}, 2025.

\bibitem{baum2021sok}
Carsten Baum, James Hsin-yu Chiang, Bernardo David, Tore~Kasper Frederiksen, and Lorenzo Gentile.
\newblock Sok: Mitigation of front-running in decentralized finance.
\newblock {\em Cryptology ePrint Archive}, 2021.

\bibitem{cachin2001secure}
Christian Cachin, Klaus Kursawe, Frank Petzold, and Victor Shoup.
\newblock Secure and efficient asynchronous broadcast protocols.
\newblock In {\em Annual Int. Cryptology Conf.}, pages 524--541. Springer, 2001.

\bibitem{cachin2022quick}
Christian Cachin, Jovana Mi{\'c}i{\'c}, and Nathalie Steinhauer.
\newblock Quick order fairness.
\newblock In {\em Int. Conf. on Financial Cryptography and Data Security (FC)}, pages 1--18. Springer, 2022.

\bibitem{alephbftrepo2026}
{Cardinal Cryptography}.
\newblock Alephbft.
\newblock \url{https://github.com/Cardinal-Cryptography/AlephBFT}, 2026.
\newblock Public code repository for AlephBFT.

\bibitem{celo}
{Celo}.
\newblock Celo: Ethereum layer 2 for payments, stablecoins and defi.
\newblock \url{https://celo.org/}, 2026.
\newblock Accessed 2026.

\bibitem{mev2023chainlink}
Chainlink.
\newblock What is maximal extractable value (mev)?
\newblock https://chain.link/education-hub/maximal-extractable-value-mev, 2023.

\bibitem{chainlink}
{Chainlink}.
\newblock Chainlink: The industry-standard oracle platform.
\newblock \url{https://chain.link/}, 2026.
\newblock Accessed 2026.

\bibitem{cheng2024shardag}
Feng Cheng, Jiang Xiao, Cunyang Liu, Shijie Zhang, Yifan Zhou, Bo~Li, Baochun Li, and Hai Jin.
\newblock Shardag: Scaling dag-based blockchains via adaptive sharding.
\newblock In {\em Int. Conf. on Data Engineering (ICDE)}, pages 2068--2081. IEEE, 2024.

\bibitem{crain2021red}
Tyler Crain, Christopher Natoli, and Vincent Gramoli.
\newblock Red belly: a secure, fair and scalable open blockchain.
\newblock In {\em Symposium on Security and Privacy (SP)}. IEEE, 2021.

\bibitem{dai2023gradeddag}
Xiaohai Dai, Zhaonan Zhang, Jiang Xiao, Jingtao Yue, Xia Xie, and Hai Jin.
\newblock Gradeddag: An asynchronous dag-based bft consensus with lower latency.
\newblock In {\em Int. Symposium on Reliable Distributed Systems (SRDS)}, pages 107--117. IEEE, 2023.

\bibitem{daian2020flash}
Philip Daian, Steven Goldfeder, Tyler Kell, Yunqi Li, Xueyuan Zhao, Iddo Bentov, Lorenz Breidenbach, and Ari Juels.
\newblock Flash boys 2.0: Frontrunning in decentralized exchanges, miner extractable value, and consensus instability.
\newblock In {\em Symposium on Security and Privacy (SP)}, pages 910--927. IEEE, 2020.

\bibitem{danezis2022narwhal}
George Danezis, Lefteris Kokoris-Kogias, Alberto Sonnino, and Alexander Spiegelman.
\newblock Narwhal and tusk: a dag-based mempool and efficient bft consensus.
\newblock In {\em European Conf. on Computer Systems (EuroSys)}, pages 34--50, 2022.

\bibitem{defillama_sui}
{DefiLlama}.
\newblock {Sui} {DEX} volume.
\newblock \url{https://defillama.com/dexs/chain/sui}, 2026.
\newblock Accessed July 2026.

\bibitem{duplyakin2019design}
Dmitry Duplyakin, Robert Ricci, Aleksander Maricq, Gary Wong, Jonathon Duerig, Eric Eide, Leigh Stoller, Mike Hibler, David Johnson, Kirk Webb, et~al.
\newblock The design and operation of $\{$CloudLab$\}$.
\newblock In {\em Annual Technical Conf. (ATC)}, pages 1--14. USENIX Association, 2019.

\bibitem{eskandari2019sok}
Shayan Eskandari, Seyedehmahsa Moosavi, and Jeremy Clark.
\newblock Sok: Transparent dishonesty: front-running attacks on blockchain.
\newblock In {\em Int. Conf. on Financial Cryptography and Data Security (FC)}, pages 170--189. Springer, 2019.

\bibitem{narwhalrepo2026}
{facebookresearch}.
\newblock Narwhal.
\newblock \url{https://github.com/facebookresearch/narwhal}, 2026.
\newblock Public code repository for Narwhal and Tusk.

\bibitem{torres2021frontrunner}
Christof Ferreira~Torres, Ramiro Camino, and Radu State.
\newblock Frontrunner {J}ones and the raiders of the dark forest: An empirical study of frontrunning on the {E}thereum blockchain.
\newblock In {\em USENIX Security Symposium}, pages 1343--1359. USENIX Association, 2021.

\bibitem{torres2024rolling}
Christof Ferreira~Torres, Albin Mamuti, Ben Weintraub, Cristina Nita-Rotaru, and Shweta Shinde.
\newblock Rolling in the shadows: Analyzing the extraction of {MEV} across layer-2 rollups.
\newblock In {\em SIGSAC Conference on Computer and Communications Security (CCS)}. ACM, 2024.

\bibitem{fischer1985impossibility}
Michael~J Fischer, Nancy~A Lynch, and Michael~S Paterson.
\newblock Impossibility of distributed consensus with one faulty process.
\newblock {\em Journal of the ACM (JACM)}, 32(2):374--382, 1985.

\bibitem{gilad2017algorand}
Yossi Gilad, Rotem Hemo, Silvio Micali, Georgios Vlachos, and Nickolai Zeldovich.
\newblock Algorand: Scaling byzantine agreements for cryptocurrencies.
\newblock In {\em Symposium on Operating Systems Principles (SOSP)}, pages 51--68. ACM, 2017.

\bibitem{giridharan2024autobahn}
Neil Giridharan, Florian Suri-Payer, Ittai Abraham, Lorenzo Alvisi, and Natacha Crooks.
\newblock Autobahn: Seamless high speed bft.
\newblock In {\em Symposium on Operating Systems Principles (SOSP)}, pages 1--23. ACM SIGOPS, 2024.

\bibitem{gorman2025vraas}
Jacob Gorman, Lucjan Hanzlik, Aniket Kate, Easwar~Vivek Mangipudi, Pratyay Mukherjee, Pratik Sarkar, and Sri~AravindaKrishnan Thyagarajan.
\newblock Vraas: Verifiable randomness as a service on blockchains.
\newblock In {\em Computer Security Foundations Symposium (CSF)}, pages 331--346. IEEE, 2025.

\bibitem{heimbach2022sok}
Lioba Heimbach and Roger Wattenhofer.
\newblock Sok: Preventing transaction reordering manipulations in decentralized finance.
\newblock In {\em Conf. on Advances in Financial Technologies (AFT)}, pages 1--14. ACM, 2022.

\bibitem{hu2026lemonshark}
Michael~Yiqing Hu, Alvin Hong~Yao Yan, Yihan Yang, Xiang Liu, and Jialin Li.
\newblock Lemonshark: Asynchronous $\{$DAG-BFT$\}$ with early finality.
\newblock In {\em Symposium on Networked Systems Design and Implementation (NSDI)}, pages 469--492. USENIX Association, 2026.

\bibitem{kang2025fairdag}
Dakai Kang, Junchao Chen, Tien Tuan~Anh Dinh, and Mohammad Sadoghi.
\newblock Fairdag: consensus fairness over multi-proposer causal design.
\newblock {\em Proceedings of the VLDB Endowment}, 19(2):265--278, 2025.

\bibitem{kavousi2025blindperm}
Alireza Kavousi, Duc~V. Le, Philipp Jovanovic, and George Danezis.
\newblock {BlindPerm}: Efficient {MEV} mitigation with an encrypted mempool and permutation.
\newblock In {\em Int. Conf. on Principles of Distributed Systems (OPODIS)}, 2025.

\bibitem{keidar2021all}
Idit Keidar, Eleftherios Kokoris-Kogias, Oded Naor, and Alexander Spiegelman.
\newblock All you need is dag.
\newblock In {\em Symposium on Principles of Distributed Computing (PODC)}, pages 165--175. ACM, 2021.

\bibitem{kelkar2022order}
Mahimna Kelkar, Soubhik Deb, and Sreeram Kannan.
\newblock Order-fair consensus in the permissionless setting.
\newblock In {\em ASIA Public-Key Cryptography Workshop}, pages 3--14. ACM, 2022.

\bibitem{kelkar2023themis}
Mahimna Kelkar, Soubhik Deb, Sishan Long, Ari Juels, and Sreeram Kannan.
\newblock Themis: Fast, strong order-fairness in byzantine consensus.
\newblock In {\em SIGSAC Conf. on Computer and Communications Security (CCS)}, pages 475--489. ACM, 2023.

\bibitem{kelkar2020order}
Mahimna Kelkar, Fan Zhang, Steven Goldfeder, and Ari Juels.
\newblock Order-fairness for byzantine consensus.
\newblock In {\em Annual Int. Cryptology Conf.}, pages 451--480. Springer, 2020.

\bibitem{kiayias2017ouroboros}
Aggelos Kiayias, Alexander Russell, Bernardo David, and Roman Oliynykov.
\newblock Ouroboros: A provably secure proof-of-stake blockchain protocol.
\newblock In {\em Annual Int. Cryptology Conf.}, pages 357--388. Springer, 2017.

\bibitem{klages2019stability}
Ariah Klages-Mundt and Andreea Minca.
\newblock (in) stability for the blockchain: Deleveraging spirals and stablecoin attacks.
\newblock {\em arXiv preprint arXiv:1906.02152}, 2019.

\bibitem{kokoris2018omniledger}
Eleftherios Kokoris-Kogias, Philipp Jovanovic, Linus Gasser, Nicolas Gailly, Ewa Syta, and Bryan Ford.
\newblock Omniledger: A secure, scale-out, decentralized ledger via sharding.
\newblock In {\em Symposium on Security and Privacy (SP)}, pages 583--598. IEEE, 2018.

\bibitem{kursawe2020wendy}
Klaus Kursawe.
\newblock Wendy, the good little fairness widget: Achieving order fairness for blockchains.
\newblock In {\em Conf. on Advances in Financial Technologies (AFT)}, pages 25--36. ACM, 2020.

\bibitem{kursawe2021wendy}
Klaus Kursawe.
\newblock Wendy grows up: More order fairness.
\newblock In {\em Int. Conf. on Financial Cryptography and Data Security (FC)}, pages 191--196. Springer, 2021.

\bibitem{lamport1978time}
Leslie Lamport.
\newblock Time, clocks, and the ordering of events in a distributed system.
\newblock {\em Communications of the ACM}, 21(7):558--565, 1978.

\bibitem{lev2020fairledger}
Kfir Lev-Ari, Alexander Spiegelman, Idit Keidar, and Dahlia Malkhi.
\newblock Fairledger: A fair blockchain protocol for financial institutions.
\newblock In {\em Int. Conf. on Principles of Distributed Systems (OPODIS)}. Schloss Dagstuhl-Leibniz-Zentrum fuer Informatik, 2019.

\bibitem{li2024sok}
Zhuolun Li and Evangelos Pournaras.
\newblock Sok: Consensus for fair message ordering.
\newblock {\em arXiv preprint arXiv:2411.09981}, 2024.

\bibitem{mahe2025order}
Erwan Mahe and Sara Tucci-Piergiovanni.
\newblock Order fairness evaluation of dag-based ledgers.
\newblock In {\em Int. Confe. on Blockchain Computing and Applications (BCCA)}, pages 106--114. IEEE, 2025.

\bibitem{malkhi2022maximal}
Dahlia Malkhi and Pawel Szalachowski.
\newblock Maximal extractable value (mev) protection on a dag.
\newblock In {\em Int. Conf. on Blockchain Economics, Security and Protocols (Tokenomics)}, pages 1--17, 2022.

\bibitem{miller2016honey}
Andrew Miller, Yu~Xia, Kyle Croman, Elaine Shi, and Dawn Song.
\newblock The honey badger of bft protocols.
\newblock In {\em Conf. on Computer and Communications Security (CCS)}, pages 31--42. ACM, 2016.

\bibitem{mirzaei2026fair}
Iliya Mirzaei and Mohammad~Javad Amiri.
\newblock Fair on the surface: Transaction-ordering bias and {MEV} in {Mysticeti} {DAG}-based {BFT} protocol.
\newblock {\em arXiv preprint arXiv:2607.13378}, 2026.

\bibitem{mirzaei2027benchmark}
Iliya Mirzaei, Zichun Cai, Chenyuan Wu, and Mohammad~Javad Amiri.
\newblock Transaction order under attack: Benchmarking {MEV} in {DAG}-based {BFT} consensus protocols.
\newblock In {\em Int. Conf. on Management of Data (SIGMOD)}. ACM, 2027.
\newblock To appear.

\bibitem{mysticetirepo2026}
{MystenLabs}.
\newblock Mysticeti.
\newblock \url{https://github.com/MystenLabs/mysticeti}, 2026.
\newblock Public code repository for Mysticeti.

\bibitem{suirepo2026}
{MystenLabs}.
\newblock Sui.
\newblock \url{https://github.com/MystenLabs/sui}, 2026.
\newblock Public code repository used for the Bullshark-based path in our evaluation.

\bibitem{nagda2026dag}
Heena Nagda, Sidharth Sankhe, Sakshi Sinha, Keon Attarha, Mohammad~Javad Amiri, and Boon~Thau Loo.
\newblock Dag of dags: Order-fairness made practical.
\newblock In {\em SIGMOD Int. Conf. on Management of Data}. ACM, 2026.

\bibitem{nagda2024rashnu}
Heena Nagda, Shubhendra~Pal Singhal, Mohammad~Javad Amiri, and Boon~Thau Loo.
\newblock Rashnu: Data-dependent order-fairness.
\newblock {\em Proceedings of the VLDB Endowment}, 17(9):2335--2348, 2024.

\bibitem{nakamoto2008bitcoin}
Satoshi Nakamoto.
\newblock Bitcoin: A peer-to-peer electronic cash system.
\newblock 2008.

\bibitem{autobahnrepo2026}
{Neil Giri}.
\newblock autobahn-artifact.
\newblock \url{https://github.com/neilgiri/autobahn-artifact}, 2026.
\newblock Public implementation repository configured for the Autobahn evaluation path; benchmark settings point to branch \texttt{autobahn}.

\bibitem{mahimahirepo2026}
{Pasindu Tennage}.
\newblock Mahi-mahi consensus.
\newblock \url{https://github.com/PasinduTennage/mahi-mahi-consensus}, 2026.
\newblock Public code repository for Mahi-Mahi.

\bibitem{pass2017hybrid}
Rafael Pass and Elaine Shi.
\newblock Hybrid consensus: Efficient consensus in the permissionless model.
\newblock In {\em Int. Symposium on Distributed Computing (DISC)}, page~6, 2017.

\bibitem{putnik2026herring}
Marko Putnik and J{\'e}r{\'e}mie Decouchant.
\newblock Herring: Parallel batch-order-fairness on dag-based blockchain consensus.
\newblock {\em arXiv preprint arXiv:2605.23648}, 2026.

\bibitem{qin2022quantifying}
Kaihua Qin, Liyi Zhou, and Arthur Gervais.
\newblock Quantifying blockchain extractable value: How dark is the forest?
\newblock In {\em Symposium on Security and Privacy (SP)}, pages 198--214. IEEE, 2022.

\bibitem{qin2021attacking}
Kaihua Qin, Liyi Zhou, Benjamin Livshits, and Arthur Gervais.
\newblock Attacking the {DeFi} ecosystem with flash loans for fun and profit.
\newblock In {\em Financial Cryptography and Data Security (FC)}, pages 3--32. Springer, 2021.

\bibitem{schneider1990implementing}
Fred~B Schneider.
\newblock Implementing fault-tolerant services using the state machine approach: A tutorial.
\newblock {\em Computing Surveys (CSUR)}, 22(4):299--319, 1990.

\bibitem{shio}
{Shio}.
\newblock {Shio}: {MEV} protection infrastructure for {Sui}.
\newblock \url{https://getshio.com}, 2024.
\newblock Accessed July 2026.

\bibitem{shrestha2024sailfish}
Nibesh Shrestha, Rohan Shrothrium, Aniket Kate, and Kartik Nayak.
\newblock Sailfish: Towards improving the latency of dag-based bft.
\newblock In {\em Symposium on Security and Privacy (SP)}, pages 21--21. IEEE, 2024.

\bibitem{shrivastav2019shoal}
Vishal Shrivastav, Asaf Valadarsky, Hitesh Ballani, Paolo Costa, Ki~Suh Lee, Han Wang, Rachit Agarwal, and Hakim Weatherspoon.
\newblock Shoal: A network architecture for disaggregated racks.
\newblock In {\em Symposium on Networked Systems Design and Implementation (NSDI)}, pages 255--270. USENIX Association, 2019.

\bibitem{singh2008bft}
Atul Singh, Tathagata Das, Petros Maniatis, Peter Druschel, and Timothy Roscoe.
\newblock Bft protocols under fire.
\newblock In {\em Symposium on Networked Systems Design and Implementation (NSDI)}, volume~8, pages 189--204. USENIX Association, 2008.

\bibitem{spiegelman2022bullshark}
Alexander Spiegelman, Neil Giridharan, Alberto Sonnino, and Lefteris Kokoris-Kogias.
\newblock Bullshark: Dag bft protocols made practical.
\newblock In {\em ACM SIGSAC Conf. on Computer and Communications Security (CCS)}, pages 2705--2718, 2022.

\bibitem{stathakopoulou2021adding}
Chrysoula Stathakopoulou, Signe R{\"u}sch, Marcus Brandenburger, and Marko Vukoli{\'c}.
\newblock Adding fairness to order: Preventing front-running attacks in bft protocols using tees.
\newblock In {\em Int. Symp on Reliable Distributed Systems (SRDS)}, pages 34--45. IEEE, 2021.

\bibitem{sui}
{Suiscan}.
\newblock Suiscan project directory.
\newblock \url{https://suiscan.xyz/mainnet/apps/directory/}, 2026.
\newblock Accessed 2026.

\bibitem{supra}
{Supra}.
\newblock Supra: A faster, better web3 experience for everyone.
\newblock \url{https://supra.com/}, 2026.
\newblock Accessed 2026.

\bibitem{wadhwa2024data}
Sarisht Wadhwa, Luca Zanolini, Aditya Asgaonkar, Francesco D'Amato, Chengrui Fang, Fan Zhang, and Kartik Nayak.
\newblock Data independent order policy enforcement: Limitations and solutions.
\newblock In {\em SIGSAC Conf. on Computer and Communications Security (CCS)}. ACM, 2024.

\bibitem{wood2014ethereum}
Gavin Wood.
\newblock Ethereum: A secure decentralised generalised transaction ledger.
\newblock {\em Ethereum project yellow paper}, 151:1--32, 2014.

\bibitem{xu2023sok}
Jiahua Xu, Krzysztof Paruch, Simon Cousaert, and Yebo Feng.
\newblock {SoK}: Decentralized exchanges ({DEX}) with automated market maker ({AMM}) protocols.
\newblock {\em ACM Computing Surveys}, 55(11):1--50, 2023.

\bibitem{yakira2021helix}
David Yakira, Avi Asayag, Gad Cohen, Ido Grayevsky, Maya Leshkowitz, Ori Rottenstreich, and Ronen Tamari.
\newblock Helix: A fair blockchain consensus protocol resistant to ordering manipulation.
\newblock {\em IEEE Transactions on Network and Service Management}, 18(2):1584--1597, 2021.

\bibitem{yang2024sok}
Sen Yang, Fan Zhang, Ken Huang, Xi~Chen, Youwei Yang, and Feng Zhu.
\newblock {SoK}: {MEV} countermeasures: Theory and practice.
\newblock In {\em Workshop on Decentralized Finance and Security (DeFi)}, 2024.

\bibitem{zhang2024no}
Jianting Zhang and Aniket Kate.
\newblock No fish is too big for flash boys! frontrunning on dag-based blockchains.
\newblock {\em Cryptology ePrint Archive}, 2024.

\bibitem{zhang2020byzantine}
Yunhao Zhang, Srinath Setty, Qi~Chen, Lidong Zhou, and Lorenzo Alvisi.
\newblock Byzantine ordered consensus without byzantine oligarchy.
\newblock In {\em Symposium on Operating Systems Design and Implementation (OSDI)}, pages 633--649. USENIX Association, 2020.

\bibitem{zhou2021high}
Liyi Zhou, Kaihua Qin, Christof~Ferreira Torres, Duc~V Le, and Arthur Gervais.
\newblock High-frequency trading on decentralized on-chain exchanges.
\newblock In {\em Symposium on Security and Privacy (SP)}, pages 428--445. IEEE, 2021.

\end{thebibliography}

\end{document}